\documentclass[journal=apchd5,manuscript=article]{achemso}

\usepackage[version=3]{mhchem} 

\usepackage{amsmath,amsfonts,amssymb}
\usepackage{graphicx}
\usepackage[colorlinks=true, allcolors=blue]{hyperref}
\usepackage{nameref}
\usepackage{physics}
\usepackage{multirow}
\usepackage{makecell}
\usepackage{siunitx}
\DeclareSIUnit{\percent}{\%}

\author{Samuele Brunetta}
\affiliation[EPFL LASPE]
{Laboratory of Advanced Semiconductors for Photonics and Electronics, Ecole Polytechnique Fédérale de Lausanne (EPFL), CH-1015 Lausanne, Switzerland}
\email{samuele.brunetta@epfl.ch}

\author{Samantha Sbarra}
\affiliation[EPFL PHOSL]
{Photonic Systems Laboratory, Ecole Polytechnique Fédérale de Lausanne (EPFL), CH-1015 Lausanne, Switzerland}
\author{Brandon Shuen Yi Loke}
\affiliation[EPFL PHOSL]
{Photonic Systems Laboratory, Ecole Polytechnique Fédérale de Lausanne (EPFL), CH-1015 Lausanne, Switzerland}
\author{Jean-François Carlin}
\affiliation[EPFL LASPE]
{Laboratory of Advanced Semiconductors for Photonics and Electronics, Ecole Polytechnique Fédérale de Lausanne (EPFL), CH-1015 Lausanne, Switzerland}
\author{Nicolas Grandjean}
\affiliation[EPFL LASPE]
{Laboratory of Advanced Semiconductors for Photonics and Electronics, Ecole Polytechnique Fédérale de Lausanne (EPFL), CH-1015 Lausanne, Switzerland}
\author{Camille-Sophie Brès}
\affiliation[EPFL PHOSL]
{Photonic Systems Laboratory, Ecole Polytechnique Fédérale de Lausanne (EPFL), CH-1015 Lausanne, Switzerland}
\author{Raphaël Butté}
\affiliation[EPFL LASPE]
{Laboratory of Advanced Semiconductors for Photonics and Electronics, Ecole Polytechnique Fédérale de Lausanne (EPFL), CH-1015 Lausanne, Switzerland}

\title{Sputtered AlN buffer layer for low-loss crystalline AlN-on-sapphire integrated photonics}

\keywords{Aluminum nitride, microring resonator, voids, photonic integrated circuits, metalorganic vapor-phase epitaxy, sputtered buffer layer}

\begin{document}


\begin{abstract}
In recent years, aluminum nitride (AlN) has emerged as an attractive material for integrated photonics due to its low propagation losses, wide transparency window, and presence of both second- and third-order optical nonlinearities. However, most of the research led on this platform has primarily focused on applications, rather than material optimization, although the latter is equally important to ensure its technological maturity. In this work, we show that voids, which are commonly found in crystalline AlN-on-sapphire epilayers, have a detrimental role in related photonic structures, as they can lead to propagation losses exceeding \SI{30}{\decibel\per\centi\meter} at 1550 nm. Their impact on light propagation is further quantified through finite-difference time-domain simulations that reveal that void-related scattering losses are strongly dependent on their size and density in the layer. As a possible solution, we demonstrate that when introducing a thin sputtered AlN buffer layer prior to initiating AlN epitaxial growth, void-free layers are obtained. They exhibit intrinsic quality factors in microring resonators as high as $2.0\times 10^6$, corresponding to propagation losses lower than \SI{0.2}{\decibel\per\centi\meter} at 1550 nm. These void-free layers are further benchmarked for high-power applications through second-harmonic and supercontinuum generation in dispersion-engineered waveguides.
Such layers are highly promising candidates for short-wavelength photonic integrated circuit applications, particularly given the strong potential of AlN for visible photonics. Given that volumetric scattering losses scale as $\lambda^{-4}$, the platform quality becomes increasingly critical in the visible and ultraviolet range, where our improved layers are expected to deliver enhanced performance.

\end{abstract}

\section{Introduction} \label{par:introduction}

In the last decade, aluminum nitride (AlN) has emerged as an attractive material for linear and nonlinear integrated photonics due to its outstanding properties. Chief among them is its wide transparency window, which ranges from \SI{210}{\nano\meter} to beyond \SI{13}{\micro\meter},\cite{morkoc_handbook_2008}, enabling a broad range of photonic applications from the ultraviolet \cite{liu_ultra-high-q_2018} to the mid-infrared\cite{dong_aluminum_2019}. In addition to a third-order optical nonlinearity mediated by the nonlinear refractive index $n_2= \SI{3.5d-19}{\square\meter\per\watt}$ of the same order as for competing platforms such as silicon nitride (Si\textsubscript{3}N\textsubscript{4}) and thin-film lithium niobate (TFLN)\cite{liu_aluminum_2023, ikeda_thermal_2008, zhu_integrated_2021}, AlN also possesses an intrinsic second-order susceptibility $\chi^{(2)}$ originating from its noncentrosymmetric wurtzite cell structure. These aspects, together with its suitability to handle high optical powers due to its relatively small thermo-optic coefficient ($ \dv*{n_{\text{op}}}{T} \simeq \SI{2.3d-5}{\per\kelvin}$ at 1550 nm)\cite{watanabe_temperature_2008} and good thermal conductivity ($\kappa = \SI{285}{\watt\per\meter\per\kelvin}$),\cite{slack_intrinsic_1987}  make it a sound choice for nonlinear optical applications \cite{honda_229_2023, bruch_17_2018, liu_integrated_2018, lu_ultraviolet_2020}.

Although the first AlN photonic integrated circuits (PICs) were fabricated from sputtered AlN layers on  SiO\textsubscript{2}-on-Si substrates\cite{pernice_second_2012}, since 2017 most of AlN PIC studies focused on AlN layers grown on \textit{c}-plane sapphire by metalorganic vapor-phase epitaxy (MOVPE)\cite{liu_aluminum_2017}. The latter show improved material quality enabling the fabrication of microring resonators (MRRs) with intrinsic quality factors ($Q_{\text{int}}$) up to \SI{3.7d6}{} at \SI{1550}{\nano\meter} (corresponding to propagation losses of \SI{0.1}{\decibel\per\centi\meter})\cite{liu_fundamental_2022}, against a $Q_{\text{int}}$ value limited to \SI{8d5}{} for sputtered AlN\cite{jung_optical_2013}. Since then, a wide range of works dealing with nonlinear and quantum optics have been reported\cite{lu_bright_2020, jang_aluminum_2023}, but few efforts have been devoted to understanding the mechanisms responsible for optical losses. This research direction could open the door to further advances with this platform, as reported successfully with Si$_{3}$N$_{4}$\cite{pfeiffer_ultra-smooth_2018, corato-zanarella_absorption_2024, liu_fabrication_2025, ji_deterministic_2025} and TFLN\cite{zhang_monolithic_2017}.

In particular, due to the heteroepitaxial nature of the growth and the strategies commonly used to produce optoelectronic-grade layers, AlN epilayers grown on sapphire often contain voids\cite{wu_recent_2021}. While such a feature can be beneficial for short-wavelength optoelectronic applications as voids filter dislocations, relieve strain to support the growth of thicker layers,\cite{tang_growth_2020, he_fast_2020} and could facilitate light extraction, they may pose some challenges in the field of integrated photonics. Indeed, voids could act as efficient scattering centers that strongly degrade optical performance, an issue that becomes all the more critical for short wavelength applications, such as chip-scale self-injection locked lasers of use for compact optical atomic clocks, and quantum sensing, since volumetric scattering losses are expected to scale as $\lambda^{-4}$\cite{weiss_splitting_1995}.

In this work, we investigate the impact of voids in AlN-on-sapphire waveguiding structures. We show that voids, commonly found in MOVPE-grown layers, can lead to propagation losses exceeding \SI{30}{\decibel\per\centi\meter} at a wavelength of \SI{1550}{\nano\meter}, as measured in both waveguides (WGs) and MRRs. Such level of losses is further supported by finite-difference time-domain (FDTD) simulations, which reveal their strong sensitivity to the size and density of voids. Moreover, simulations indicate that relatively small voids, which have a negligible impact at telecom wavelengths, may play an important role in propagation losses at shorter wavelengths, highlighting voids as a potential limiting factor in current state-of-the art visible AlN photonics.   Finally, we experimentally show that ``hybrid" AlN layers, consisting of MOVPE-grown AlN on a thin sputtered AlN buffer on sapphire, being void-free, offer a powerful solution to this problem. With such layers, we achieve integrated waveguiding structures at \SI{1550}{\nano\meter} with low propagation loss  $<$\SI{0.2}{\decibel\per\centi\meter}, and we confirm their suitability for nonlinear photonic applications with the demonstration of second-harmonic generation (SHG) and supercontinuum generation (SCG) in dispersion-engineered WGs.

\section{Results and discussion} \label{par:results_and_disc}
\subsection{Material characterization} \label{par:material_char}

In this work, we studied three different types of AlN epilayers grown on 2-inch $c$-plane sapphire substrates. Sample A was grown with an Aixtron 200/4 RF-S horizontal reactor, using a full-MOVPE process, which includes a thin AlN nucleation layer grown at lower temperature to facilitate the start of the heteroepitaxial growth process. Sample B is a \SI{1.0}{\micro\meter}-thick commercially available epilayer purchased from DOWA Electronics Materials Co. Ltd. Sample C relies on an in-house hybrid process, which consists in growing the main AlN layer by MOVPE on top of a 50-nm-thick sputter-deposited AlN buffer layer on \textit{c}-plane sapphire, provided by Evatec AG. The main characterization parameters are shown in Table~\ref{tbl:layer_characterization}, while additional details and data are reported in the Supporting Information (SI) Section~S1. The three layers share a comparable value of the root mean square (RMS) surface roughness, such as deduced from atomic force microscopy (AFM). However, x-ray diffraction (XRD) data indicate that sample A has both a poorer crystalline quality and a higher degree of off-plane crystal tilt and in-plane twist compared to samples B and C, as shown by the full width at half maximum (FWHM) of the symmetric $\mathrm{(0002)}$ peak and that of the asymmetric $\mathrm{(21\bar{3}1)}$ peak (Table~\ref{tbl:layer_characterization}), respectively.

\begin{table}
  \caption{Material characterization data of the three AlN-on-sapphire epilayers investigated in this work.}
  \label{tbl:layer_characterization}
  \small
  \begin{tabular}{lccc}
  \hline
    & Sample A  & Sample B  & Sample C \\
    \hline
    Sample details & \makecell{Full-MOVPE} & \makecell{Commercially \\ available} & \makecell{Hybrid\textsuperscript{\emph{a}}}  \\ \hline
    AlN thickness\textsuperscript{\emph{b}} (\SI{}{\micro\meter}) & \SI{1.1}{}   & \SI{1.0}{} & \SI{1.1}{} \\
    RMS surface roughness\textsuperscript{\emph{c}} ($\SI{}{\nano\meter}$) & \SI{0.17}{} & \SI{0.11}{} & \SI{0.09}{} \\
    XRD $\mathrm{(0002)}$ peak FWHM & \ang{0.151} & \ang{0.029} & \ang{0.023} \\
    XRD $\mathrm{(21\bar{3}1)}$ peak FWHM & \ang{0.474} & \ang{0.410} & \ang{0.441} \\
    \hline
    Void density $N$ ($\SI{}{\per\square\micro\meter}$) & $123\pm 11$ & $41\pm 6$ & None \\
    Void height $H$ ($\SI{}{\nano\meter}$) & 40-300 & 20-80 & None \\
    Void diameter $d$, deduced from SEM ($\SI{}{\nano\meter}$) & 7-13 & 4-8 & None \\
    Void diameter $d$, deduced from AFM ($\SI{}{\nano\meter}$) & 39-53 & 39-51 & None \\
    \hline
  \end{tabular} \\
   \textsuperscript{\emph{a}} MOVPE regrowth on a sputtered buffer layer.
  \textsuperscript{\emph{b}} Measured at the wafer center.
  \textsuperscript{\emph{c}} Over a $2\times \SI{2}{\square\micro\meter}$ area.
\end{table}

\begin{figure}
   \begin{center}
   \includegraphics[width=1.0\textwidth]{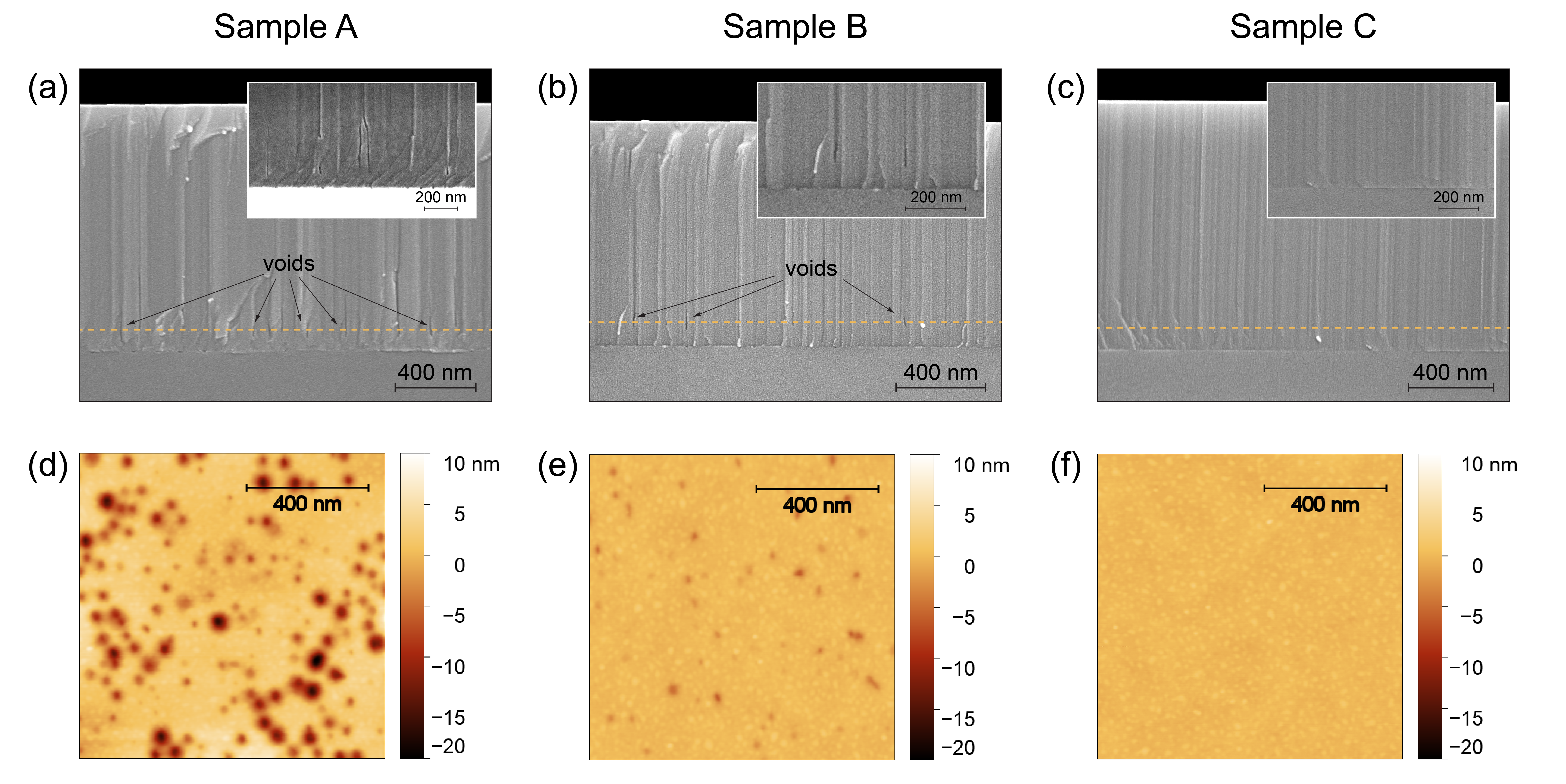}
   \end{center}
   \caption{(a)-(c) Cross-section SEM images of the epilayers A, B and C, respectively, with a high magnification inset for the AlN-sapphire interface region. Samples are coated with a thin metal layer for charge-dissipation purposes.  (d)-(f) Surface AFM scans of the epilayers A, B and C, respectively, after thinning them down by ICP dry etching to a thickness of \SI{110}{\nano\meter} (horizontal dashed orange lines in the SEM pictures).}
  \label{fig:SEM_AFM}
   \end{figure} 

In order to further probe the material quality of these samples, we performed cross-section scanning electron microscopy (SEM) imaging on the three epilayers. The SEM images shown in Figures~\ref{fig:SEM_AFM}(a)-\ref{fig:SEM_AFM}(c) reveal that, in addition to vertical striations resulting from the cleavage process, samples A and B both contain vertically-elongated air voids within the AlN layer near the interface with the sapphire substrate. No such voids were observed on sample C. This feature is accounted for by the fact that the full MOVPE recipe used for growing sample A was optimized to produce Al-polar crack-free AlN layers, as commonly done to obtain device-grade short-wavelength III-nitride Al-rich optoelectronic devices.\cite{tang_growth_2020} This recipe starts with the deposition of a low-temperature AlN nucleation layer that promotes the Al-polarity of the growing surface, followed by high-temperature AlN growth. The growth parameters during the early stage of the high-temperature growth are tuned to induce three-dimensional (3D) growth, to avoid the appearance of cracks and improve the crystalline quality. Therefore, the voids in sample A are certainly formed during the coalescence of the 3D layer. Unlike sample A, the use of a thin sputter-deposited AlN buffer layer prior to initiating MOVPE growth for sample C did not require any 3D-growth stage to obtain crack-free epilayers, hence leading to void-free layers.

Since voids are generally elongated in the vertical direction, we will refer to their in-plane size, usually in the few tens of nanometer range, as their diameter ($d$), while their vertical size, whose value can exceed a few hundreds of nanometers,\cite{he_fast_2020} will be referred to as their height ($H$). By analyzing multiple cross-section SEM pictures, we measured $H$ values in the range of 40-300~nm and 20-80~nm in samples A and B, respectively. $d$ values were found to lie in the range of 7-13~nm and 4-8~nm in samples A and B, respectively. However, these values are likely underestimated due to the use of a thin metal coating for mitigating charge accumulation during imaging and the inherent difficulty of accurately measuring irregularly shaped narrow features in cross-section SEM images.

In order to extract the areal density of voids $N$ in the films, the three AlN layers were etched by means of inductively coupled plasma (ICP) dry etching down to a thickness of \SI{110+-9}{\nano\meter} (horizontal orange dashed line in Figures~\ref{fig:SEM_AFM}(a)-\ref{fig:SEM_AFM}(c)), as most of the voids we observe lie at this level. AFM images of the etched surfaces are shown in Figures~\ref{fig:SEM_AFM}(d)-\ref{fig:SEM_AFM}(f), from which we could extract a void density $N$ of $\SI{123+-11}{\per\square\micro\meter}$ and $\SI{41+-6}{\per\square\micro\meter}$ in samples A and B, respectively, while no voids were found in sample C. Although the ICP etching process is likely to alter the geometry of the voids, through an increase in their diameter, AFM images, shown in Figures~\ref{fig:SEM_AFM}(d) and ~\ref{fig:SEM_AFM}(e), are consistent with cross-sectional SEM results, as the voids in sample A appear larger than in sample B. Neglecting the deepest and widest features visible in Figure~\ref{fig:SEM_AFM}(d), which likely correspond to the tallest initial voids that were thus exposed to a longer etching process, we measured void diameters lying in the 39-53~nm and the 39-51~nm  range for samples A and B, respectively. Considering that these values are most certainly overestimated due to the etching process, and that those extracted from SEM images are probably underestimated, we conclude that the actual void diameter $d$ likely lies in between these two estimates. A summary of the extracted void density and size for both samples is provided in the bottom part of Table~\ref{tbl:layer_characterization}. 

  \begin{figure}
   \begin{center}
   \includegraphics[width=1.0\textwidth]{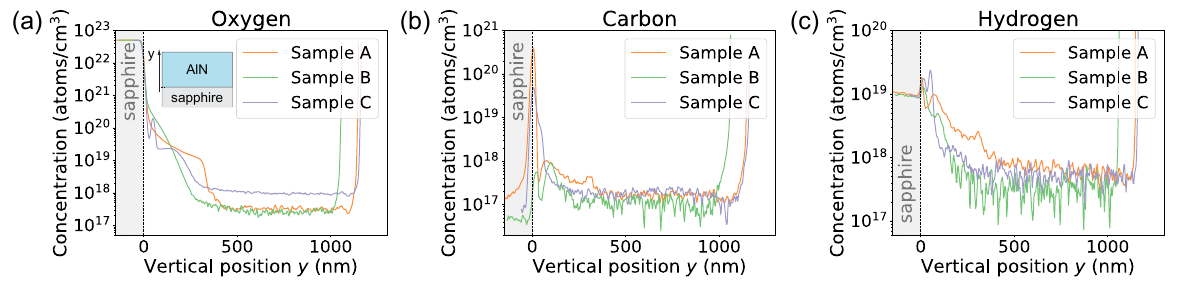}
   \end{center}
   \caption{(a) to (c) SIMS concentration profiles of the main impurity species for the three AlN-on-sapphire epilayers under investigation. The higher hydrogen concentration level measured in sample A is consistent with the presence of voids in the first \SI{400}{nm} of the layer.}
  \label{fig:SIMS}
   \end{figure} 

With the purpose of probing the commonly found impurities in MOVPE-grown AlN epilayers, secondary-ion mass spectrometry (SIMS) was performed on the three layers at EAG Laboratories. In Figure~\ref{fig:SIMS}, the impurity concentration profiles of oxygen (O), carbon (C) and hydrogen (H) are shown as a function of vertical position ($y$) in the corresponding epilayer, where $y=0$ coincides with the AlN/sapphire interface. As can be noticed in Figure~\ref{fig:SIMS}(c), the concentration of hydrogen for $y$ values comprised between 40 and \SI{400}{\nano\meter} is higher in sample A than in the void-free sample C. Since this thickness range is in agreement with the position of the voids deduced from SEM images, the higher hydrogen content is likely related to the presence of saturated dangling bonds located at the AlN/void interfaces. For the same thickness range, oxygen and carbon atoms are also more abundant in sample A than in sample C, although sample C seems to have a slightly higher oxygen background in the top part of the layer. Somewhat similar features can be observed when comparing samples B and C, though the effect is far less pronounced due to the reduced size and the lower density of voids found in sample B. These signatures suggest that a higher concentration of impurities is found in the regions where voids are present, which is in agreement with other results from the literature\cite{nagata_origin_2019}.  

\subsection{Measurement of propagation losses} \label{par:propagation_losses}

\begin{figure}
   \begin{center}
   \includegraphics[width=1.0\textwidth]{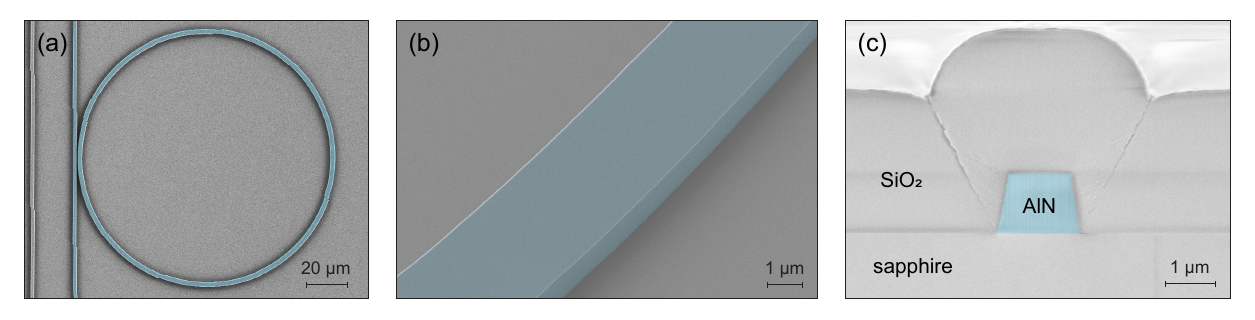}
   \end{center}
   \caption{(a) False-color top view SEM image of an AlN MRR of \SI{60}{\micro\meter} radius prior to SiO\textsubscript{2} cladding deposition. (b) Enlarged bird's eye view of an MRR showing a smooth sidewall. (c) False-color cross-section SEM image of a bus WG facet in sample C, revealing a sidewall angle of about \ang{80}.}
  \label{fig:SEM_fab}
   \end{figure}

In order to quantify propagation losses occurring in samples A to C, WGs and MRRs were fabricated using an electron-beam lithography (EBL) step followed by full-depth AlN ICP dry etching and plasma-enhanced chemical vapor deposition (PECVD) of an SiO\textsubscript{2} top cladding layer. Additional details on the fabrication process are reported in the \nameref{par:method} section of the main text. A top-view false-color SEM image of a MRR with \SI{60}{\micro\meter} radius prior to cladding deposition is presented in Figure~\ref{fig:SEM_fab}(a), while Figure~\ref{fig:SEM_fab}(b) features a zoomed-in tilted-view image showing smooth AlN sidewalls. Finally, a cross-section SEM view of a WG facet is shown in Figure~\ref{fig:SEM_fab}(c), from which we measure an AlN sidewall angle of \ang{80}, which is consistent with the values reported in the literature\cite{liu_aluminum_2017, liu_ultra-high-q_2018}. 

For WGs, propagation losses at \SI{1550}{\nano\meter} were obtained using the cut-back method, i.e., by measuring the light power transmission for different waveguide lengths and fitting the data in dB scale with a linear model. For MRRs, light from a tunable laser emitting in the C-band was first coupled to bus WGs of upper width = \SI{1.2}{\micro\meter} through a microlensed fiber, leading to evanescent coupling to rings of different widths and varying bus-to-ring gaps. The signal outcoupled from the bus WGs was collected either by an aspheric lens or a microlensed fiber. Transmission data were fitted with a Lorentzian function and the loaded quality factor ($Q_{\text{L}}$) was deduced by dividing the wavelength at the resonance ($\lambda_0$) by the FWHM of the resonance peak. The rings operated in the undercoupled regime, allowing the calculation of the intrinsic $Q$ factor $Q_{\text{int}}$ from the relationship:\cite{liu_ultra-high-q_2018} 
\begin{equation}
    Q_{\text{int}}=\frac{2Q_{\text{L}}}{1+\sqrt{T_0}},
    \label{Eq:Qint}
\end{equation}
where $T_0$ is the normalized on-resonance transmission. Finally, we extracted the propagation loss coefficient $\alpha$ from:\cite{pernice_high_2012}
\begin{equation}
    \alpha=\frac{10 \log_{10} (e) \cdot 2\pi n_\text{g}}{\lambda_0 Q_{\text{int}}},
    \label{Eq:alpha}
\end{equation}
where $n_\text{g}$ is the group index of the resonant mode under consideration, obtained from finite-difference-eigenmode simulations performed with \textit{Ansys Lumerical}.

\begin{figure}
   \begin{center}
   \includegraphics[width=1.0\textwidth]{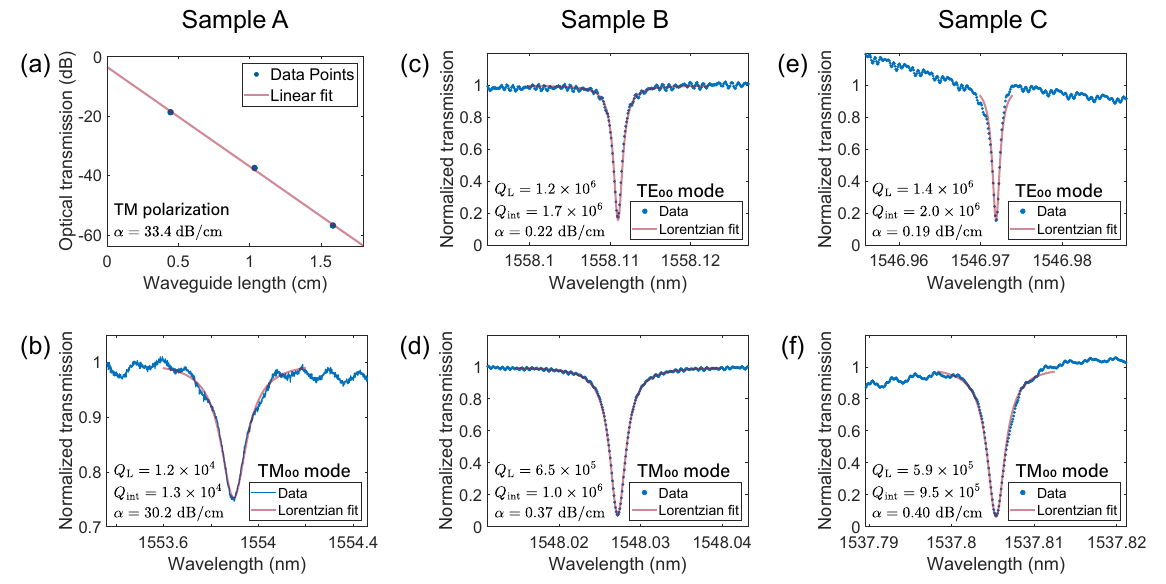}
   \end{center}
   \caption{(a) Optical transmission in \SI{1.4}{\micro\meter}-wide WGs fabricated from sample A, characterized at 1550 nm in TM polarization. (b) Transmission spectrum of a TM\textsubscript{00} resonance in a \SI{2.3}{\micro\meter}-wide MRR from sample A, showing optical losses around \SI{30}{\decibel\per\cm}. (c) to (f) Characterization of the fundamental TE and TM resonances in the C-band for \SI{1.8}{\micro\meter}-wide MRRs fabricated from samples B and C. Rings have a radius of \SI{60}{\micro\meter} and measurements were done in the undercoupled regime.}
  \label{fig:ring_Q}
   \end{figure} 

Optical performance results obtained on waveguiding structures fabricated from the three types of epilayers are shown in Figure~\ref{fig:ring_Q}. Propagation losses in \SI{1.4}{\micro\meter}-wide WGs from sample A are as high as \SI{33.4}{\decibel\per\centi\meter}, as inferred from the slope of the linear interpolation shown in Figure~\ref{fig:ring_Q}(a). Similar values are obtained for \SI{2.3}{\micro\meter}-wide MRRs fabricated from the same type of epilayer, which yielded a $Q_{\text{int}}$ value of $\SI{1.3d4}{}$ for the TM\textsubscript{00} mode, corresponding to $\alpha=\SI{30.2}{\decibel\per\centi\meter}$ (Figure~\ref{fig:ring_Q}(b)). Propagation losses in TE polarization resulted to be even higher, by $\sim$\SI{10}{\decibel\per\centi\meter} above what was measured for TM polarization. On the contrary, sample B (which contains fewer and smaller voids than sample A) and sample C (which is void-free) both yield much lower values of propagation losses, as shown in the analysis of fundamental TE and TM modes for \SI{1.8}{\micro\meter}-wide MRRs (Figures~\ref{fig:ring_Q}(c)-\ref{fig:ring_Q}(f)). More specifically, $Q_{\text{int}}$ can be as high as $\SI{2d6}{}$ for the TE\textsubscript{00} mode, (corresponding to losses $< \SI{0.2}{\decibel\per\centi\meter}$), and $\SI{1d6}{}$ for the TM\textsubscript{00} mode (corresponding to losses $<\SI{0.4}{\decibel\per\centi\meter}$). 

\begin{table}
\small
\caption{Average $Q_{\text{int}}$ values and propagation losses measured at \SI{1550}{\nano\meter} in MRRs fabricated from the three epilayers under consideration. Data are averaged over a minimum of 5 resonances per dataset.}
  \label{tbl:Q_factors}
\begin{tabular}{llccc}
\hline
                      &       & Sample A & Sample B & Sample C \\ \hline
\multirow{2}{*}{TE\textsubscript{00}} & $Q_{\text{int}}$  & Not available        & \SI{1.3+-0.3d6}{}        & \SI{1.4+-0.2d6}{}        \\
                      & $\alpha$ (\SI{}{\decibel\per\centi\meter}) & $\sim \SI{41}{}$        & \num{0.29+-0.06}        & \num{0.27+-0.04}        \\ \hline
\multirow{2}{*}{TM\textsubscript{00}} & $Q_{\text{int}}$  & \SI{1.3+-0.2d4}{}        & \SI{8.5+-1.2d5}{}        & \SI{8.7+-1.0d5}{}        \\
                      & $\alpha$ (\SI{}{\decibel\per\centi\meter}) & \num{31+-4}       & \num{0.46+-0.08}      & \num{0.44+-0.06}    
                      \\ \hline
\end{tabular}
\end{table}

\begin{figure} [ht]
   \begin{center}
   \includegraphics[width=0.45\textwidth]{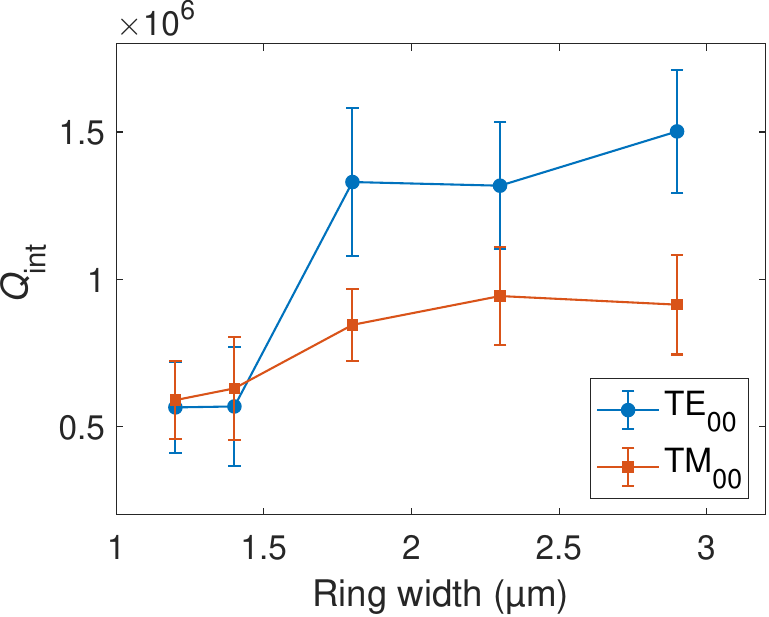}
   \end{center}
   \caption{Experimentally measured mean $Q_\text{{int}}$ value of the fundamental TE and TM resonances at $\lambda\simeq\SI{1550}{\nano\meter}$ in MRRs fabricated from sample B. The error bars correspond to the standard deviation for each MRR set. Microrings have a radius of \SI{60}{\micro\meter}, an upper width that is varied from \SI{1.2} to \SI{2.9}{\micro\meter}, and a bus WG to ring resonator coupling gap chosen to ensure that MRRs operate in the undercoupled regime.}
  \label{fig:Qint_width}
   \end{figure}

To better compare the three samples, several undercoupled ring resonances were analyzed, and the average values of $Q_{\text{int}}$ and $\alpha$, determined using Eqs.~\eqref{Eq:Qint} and \eqref{Eq:alpha}, respectively,  are reported in Table~\ref{tbl:Q_factors}, where errors represent the standard deviation of the resonance dataset. As can be noticed, a decrease in the density and the size of the voids such as those present in samples A and B is sufficient to improve $Q_{\text{int}}$ values by more than 2 orders of magnitude at telecom wavelengths. However, propagation losses reach the same magnitude for samples B and C, despite the complete absence of voids in the latter. 
We hypothesize that this limitation in both samples is rather due to optical loss channels introduced  by our fabrication procedure. 
Our assumption is further confirmed when considering the data shown in Figure~\ref{fig:Qint_width}, for which $Q_{\text{int}}$ values obtained for different ring widths in sample B are displayed. Indeed, $Q_\text{{int}}$ is smaller for narrower ring WG widths because of the larger weight of sidewall scattering, while for rings wider than \SI{1.8}{\micro\meter}, $Q_\text{{int}}$ does not significantly increase with the ring width anymore. In particular, for the TE\textsubscript{00} mode we did not obtain $Q_\text{{int}}$ values higher than \SI{1.8d6}{} in sample B. Such a value remains smaller than the $Q_\text{{int}}$ value of \SI{2.8d6}{}, reported by Y. Sun \textit{et al.} on MRRs fabricated on AlN epilayers from the same supplier, suggesting that our current results are limited by fabrication-related features.\cite{sun_ultrahigh_2019} 

The explanation accounting for our lower quality factors compared to those obtained by Y. Sun \textit{et al.}\cite{sun_ultrahigh_2019} is probably multi-factorial. First, we relied on silane-based PECVD for the SiO\textsubscript{2} cladding, which is known to leave air pockets when narrow trenches, such as those present in the case of WG-resonator gaps, are filled, as documented in SI Sec.~S2. This aspect plays an important role in MRR losses, as the vertically low lying air pocket can perturb the WG bus-microring coupling and be a non-negligible source of light scattering. In this regard, tetraethyl-orthosilicate-based PECVD, which was the cladding technique used in Ref.~\citenum{sun_ultrahigh_2019}, is expected to provide better optical results due to its superior trench refilling capabilities.\cite{chang_trench_2004} Furthermore, high-temperature annealing could be implemented in future works to further decrease optical losses occurring in the SiO\textsubscript{2} cladding layer, due to a reduction in O-H bonds\cite{henry_low_1987} that are known to be a source of light attenuation at telecom wavelengths. Lastly, the dry etching parameters could be further optimized to achieve smoother sidewalls, although we do not expect sidewall roughness to be the main source of losses in our wider rings. Indeed, based on the analytical approach proposed in Ref.~\citenum{liu_ultra-high-q_2018},  roughness-induced scattering losses are expected to scale proportionally with the normalized average field intensity at the sidewalls ($I_\mathrm{sw}$). From the FDE simulations shown in SI Sec.~S3, $I_\mathrm{sw}$ is found to decrease by a factor of two in a MRR when its width increases from \SI{1.8}{} to \SI{2.9}{\micro\meter}, leading to a corresponding twofold reduction in scattering losses. Since the quality factors shown in Figure~\ref{fig:Qint_width} do not follow this trend, we can infer that propagation losses in our samples are not limited by sidewall roughness alone, for both TE and TM polarization.

In any case, regardless of these fabrication aspects, 
we can still conclude that when  epilayers contain voids of non-negligible height (over \SI{200}{\nano\meter}) and density (over \SI{100}{\per\square\micro\meter}), they can lead to significant propagation losses at \SI{1550}{\nano\meter} exceeding \SI{30}{\decibel\per\centi\meter}, as observed in sample A.

\subsection{Simulation of void-induced scattering losses} \label{par:FDTD}

\begin{figure}
   \begin{center}
   \includegraphics[width=1.0\textwidth]{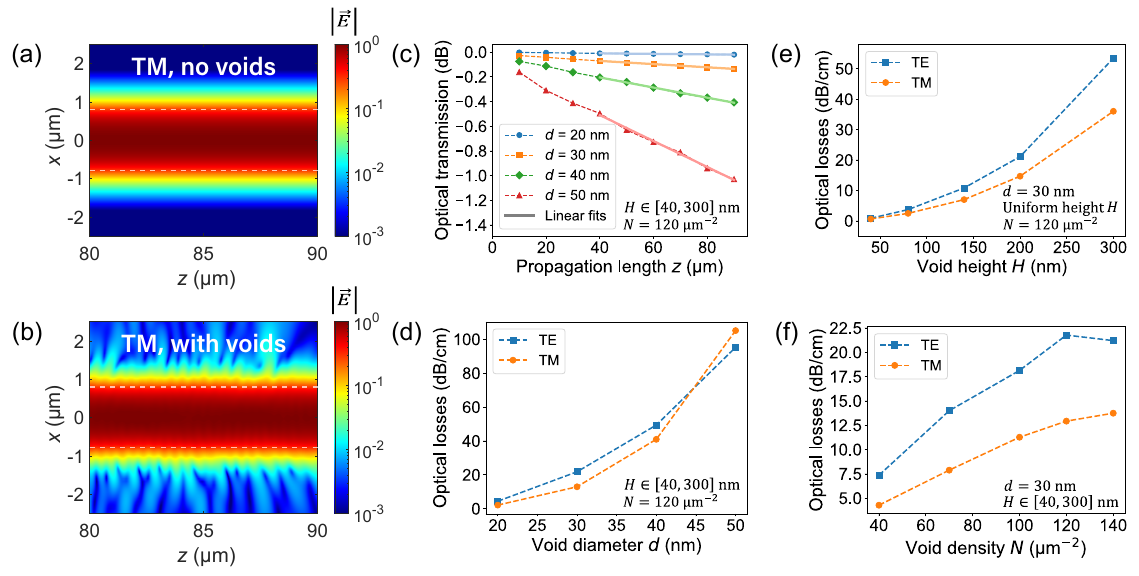}
   \end{center}
   \caption{FDTD simulations of light transmission at \SI{1550}{\nano\meter} for AlN straight WGs with different void geometries (height: \SI{1.1}{\micro\meter}, upper width: \SI{1.4}{\micro\meter}). (a) and (b) 2D top-view normalized field $|\Vec{E}|$ map at WG mid-height in the absence and in the presence of voids, respectively, showing the signature of scattering induced by the latter. The WG geometry is delimited within the two dashed lines (c) Power transmission as a function of propagation length for different void diameters $d$ at a fixed void density $N=\SI{120}{\per\square\micro\meter}$, in TM polarization. Data in decibels are fitted with a straight line to extract propagation losses. (d) Simulated optical losses as a function of $d$ for both polarizations. (e) Simulated optical losses as a function of void height $H$ for $d=\SI{30}{\nano\meter}$. (f) Simulated optical losses as a function of void density $N$ for $d=\SI{30}{\nano\meter}$.}
  \label{fig:FDTD_results}
   \end{figure} 
   
In order to confirm that voids are responsible for volumetric scattering and can be a source of significant propagation losses in fully MOVPE-grown AlN photonic structures, we performed FDTD simulations using \textit{Ansys Lumerical}\footnote{\textit{Ansys Lumerical} FDTD v. 8.30.3536 from Lumerical Inc.}. Specifically, we investigated propagation losses occurring in the telecom C-band for the fundamental TE and TM modes of an AlN-on-sapphire WG (height: \SI{1.1}{\micro\meter}, upper width: \SI{1.4}{\micro\meter})  for a total length $z$ of \SI{90}{\micro\meter}, in the absence and in the presence of voids. For the sake of simplicity, voids were modeled as square parallelepipeds, randomly generated inside the AlN WG, with sizes in agreement with the experimentally measured values for sample A, as reported in Table~\ref{tbl:layer_characterization}. Hence, the height $H$ is randomly distributed in the 40-300~nm range, while the bottom coordinate $y_\mathrm{min}$ lies in the 40-200~nm range, the areal density is fixed to  $N=\SI{120}{\per\square\micro\meter}$,  and we consider different diameter $d$ values. Perfectly-matched-layer boundary conditions were adopted in order to absorb the scattered power reaching the edges of the simulation window. A fine mesh of \SI{10}{\nano\meter} was used in the bottom part of the WG, where voids are present, while coarser meshes were used in the rest of the WG region and in the remaining parts of the simulation domain. Additional details on the simulation settings and the convergence can be found in the SI Sec.~S4.

In Figures~\ref{fig:FDTD_results}(a) and \ref{fig:FDTD_results}(b), we show a two-dimensional (2D) top-view map of the electric field modulus $|\Vec{E}|$ at WG mid-height in the absence and in the presence of voids, respectively. While the optical mode appears stable in the absence of defects, a non-negligible amount of power escapes from the WG due to scattering when voids are present. 

We monitored the optical transmission every \SI{10}{\micro\meter} along the propagation direction $z$ for different values of the void diameter $d$. Optical losses were extracted from the linear fit of the slope in dB scale, where the first three data points were excluded to minimize their misestimate because of a potentially higher collection of the scattered field in the first power monitors. As an illustrative example, we report in Figure~\ref{fig:FDTD_results}(c) the results obtained for TM polarization for which a strong dependence of transmission on $d$ is observed. The optical loss values obtained using this method for both TE and TM input polarizations are shown in Figure~\ref{fig:FDTD_results}(d). With the exception of the data points computed at $d=\SI{50}{\nano\meter}$, TE loss values are larger than TM ones, in agreement with the experimental results obtained on sample A (cf. Table~\ref{tbl:Q_factors}). We see that for a $d$ value of $\SI{30}{\nm}$, a reasonable assumption given the two estimates of the void diameter deduced from cross-section SEM and top-view AFM images, we derive losses of \SI{21.8+-0.4}{\decibel\per\centi\meter} and \SI{12.9+-0.1}{\decibel\per\centi\meter} for TE and TM polarizations, respectively. Although they are on the same order of magnitude, the simulated values are smaller than experimentally measured losses of $\simeq \SI{41}{\decibel\per\centi\meter}$ (TE) and \SI{31+-4}{\decibel\per\centi\meter} (TM) in sample A. However, given the strong dependence of the computed losses on the void size, such a difference could stem from an experimentally underestimated value of the size of the voids and their density. We also recall that higher concentrations of O, C and H impurities are found in the void region, as shown in Figure~\ref{fig:SIMS}. Therefore, part of the losses we observed in sample A could also be due to vibrational modes of impurity-related bonds, similarly to the losses induced by N-H and Si-H bonds in silicon nitride WGs\cite{douglas_effect_2016, mao_low_2008}.

Keeping $d=\SI{30}{\nm}$ and $N=\SI{120}{\per\square\micro\meter}$ as reference values, we then assigned to all voids a uniform height value to study its impact on propagation losses. Thus, as can be seen in Figure~\ref{fig:FDTD_results}(e), height seems to play a role as important as the diameter in the magnitude of those losses. Finally, simulations carried out for different void densities $N$ are shown in Figure~\ref{fig:FDTD_results}(f),  which also indicate an increase in the propagation losses with $N$, as could be anticipated.

\begin{figure}
   \begin{center}
   \includegraphics[width=1.0\textwidth]{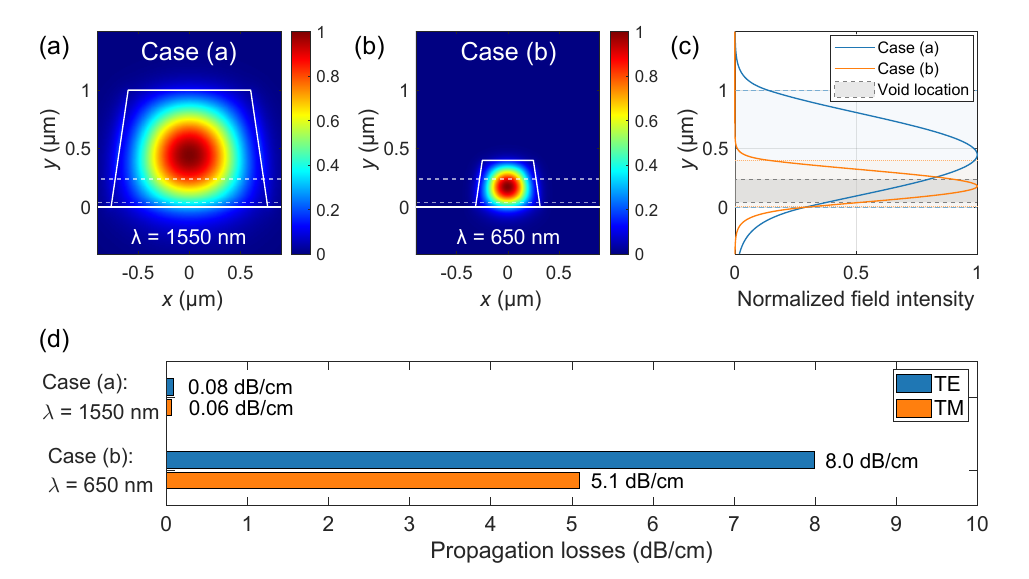}
   \end{center}
   \caption{Simulated TE\textsubscript{00} normalized field intensity ($|\vec{E}|^2$) maps computed for two different cross-section WG geometries: (a) at $\lambda=$ \SI{1550}{\nano\meter} in a $\SI{1.2}{}\times\SI{1.0}{\micro\meter}^{2}$ WG and (b) at $\lambda=$ \SI{650}{\nano\meter} in a $\SI{0.5}{}\times\SI{0.4}{\micro\meter}^{2}$ WG, where in each case the WG geometry has been chosen such that the devices operate in the single-mode regime for TE polarization. (c) Field intensity profile at the WG center for field intensity maps (a) and (b), showing a larger overlap of the field intensity distribution with the void region for case (b). (d) Losses extracted from FDTD simulations for both cases, in presence of ``small" voids in the layer.}
  \label{fig:wvl_profile}
\end{figure}

Given the remarkable difference in void density and most importantly in size between samples A and B, it is not surprising that we could not observe the impact of voids in sample B from our measurements led at \SI{1550}{\nano\meter}. However, such small voids are likely to play a non-negligible role when aiming for short-wavelength applications, e.g., in the visible and in the UV range, because the weight of volumetric scattering losses is expected to scale as $\lambda^{-4}$. Moreover, in order for a WG to remain in the single-mode regime at shorter wavelength, thinner layers are required. In such a case, as shown in Figures~\ref{fig:wvl_profile}(a) to \ref{fig:wvl_profile}(c), most of the optical mode power will reside in the region occupied by voids, leading to enhanced scattering losses compared to the situation experienced by single-mode WGs in the telecom C-band.  

To verify these hypotheses, we carried out an additional set of FDTD simulations  considering smaller voids, similar to those observed in sample B. Voids were randomly generated in the waveguide with the following parameters: $d=\SI{20}{\nano\meter}$, $H=\SI{40}{\nano\meter}$, $N=\SI{40}{\per\square\micro\meter}$, and $y_\mathrm{min}$ in the 40-200~nm range. Two cases were analyzed: a $\SI{1.2}{}\times\SI{1.0}{\micro\meter}^{2}$ WG pumped at \SI{1550}{\nano\meter} (case (a)), and a $\SI{0.5}{}\times\SI{0.4}{\micro\meter}^{2}$ WG pumped at \SI{650}{\nano\meter} (case (b)). Results are reported in Figure~\ref{fig:wvl_profile}(d), where we can observe how small voids are only responsible for negligible propagation losses at \SI{1550}{\nano\meter}, lower than \SI{0.08}{\decibel\per\centi\meter}. This explains why we could not detect significant differences in propagation losses between sample B and the void-free sample C, as losses in our structures are likely dominated by fabrication-related aspects. However, the same void distribution is expected to cause significant propagation losses at \SI{650}{\nano\meter}, namely \SI{8.0}{\decibel\per\centi\meter} in TE polarization, and \SI{5.1}{\decibel\per\centi\meter} in TM, due to a combination of increased scattering at shorter wavelengths and a higher portion of the modal field interacting with voids in this second case (as shown in Figure~\ref{fig:wvl_profile}(c)). The exact values obtained from these sets of simulations on smaller voids should be regarded as a semi-quantitative indication, since the results depend on a wide range of void parameters that are difficult to determine accurately. Nevertheless, they are useful for illustrating how even voids with a negligible impact at telecom wavelengths can become a major source of propagation losses within the visible range.

For all these reasons, the presence of voids, even of small size, is clearly detrimental for PIC applications as they can act as a major source of losses in AlN photonic structures intended for short-wavelength operation. Moreover, the losses of 5-\SI{8}{\decibel\per\centi\meter} we predicted from simulations at \SI{650}{\nano\meter} are consistent with some of the values reported in literature for AlN microring resonators operating in the visible\cite{shin_demonstration_2021}, indicating that voids could likely be one of the limiting factors in current state-of-the art AlN photonics in the visible.

Therefore, optimizing the growth process to reduce or even suppress their presence can be an effective strategy for improving the optical performance of AlN-on-sapphire PICs in this wavelength range. In this regard, being entirely void-free, the present hybrid AlN epilayers grown on a sputtered AlN buffer emerge as a promising solution for visible and UV applications with the potential to overcome current state-of-the-art values for this platform\cite{liu_ultra-high-q_2018}. 

\subsection{Nonlinear measurements on hybrid AlN-on-sapphire layers}\label{par:SCG}

\begin{figure}
   \begin{center}
   \includegraphics[width=1.0\textwidth]{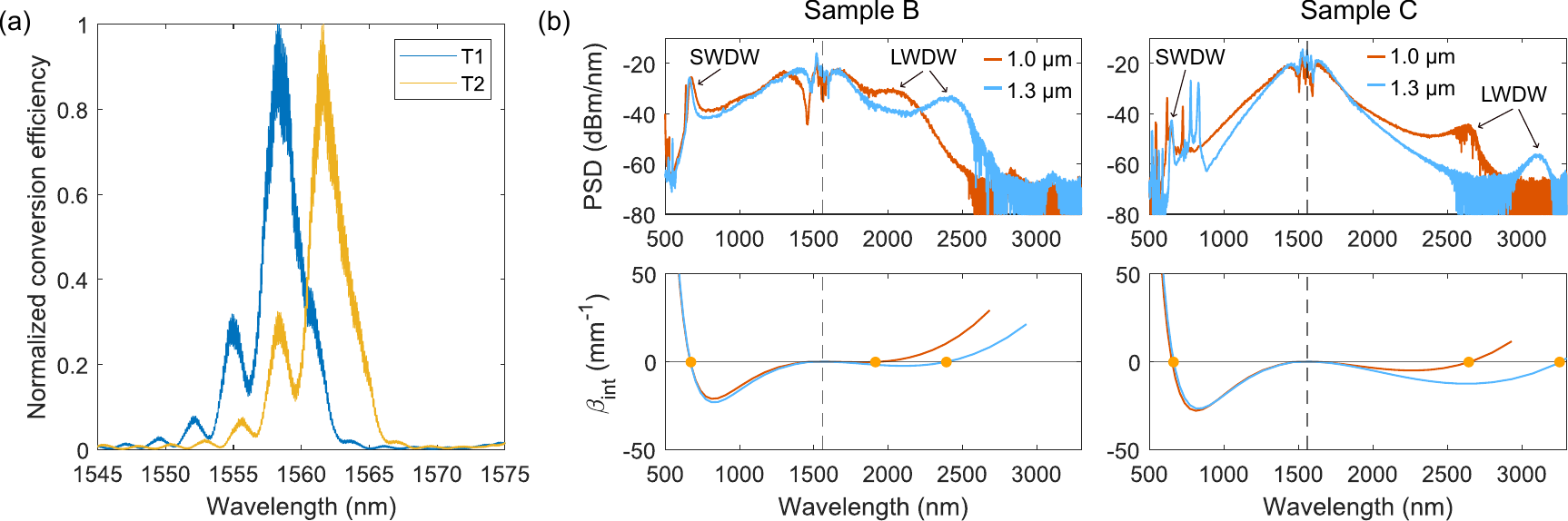}
   \end{center}
   \caption{(a) Normalized conversion efficiency of the SH power generated from a WG on sample C while sweeping the pump wavelength in the telecom range. The experiment is run at two different temperatures (T1 = 20$^\circ$C, T2 = 83$^\circ$C). (b) Comparison between the SC spectra generated by 1~µm- and 1.3~µm-wide WGs fabricated on sample B (top left) and  C (top right), respectively. The TM-polarized pump wavelength is set at 1.56~µm (dashed line) and the positions of the SWDW and LWDW are indicated. In the bottom panels the corresponding integrated dispersions ($\beta_{\text{int}}$) are plotted. The phase-matching condition for the SWDW and the LWDW is indicated by yellow dots.}
  \label{fig:Nonlinear}
\end{figure} 

To illustrate the potential of the hybrid sample C for nonlinear optical applications, we investigated the on-chip SHG and SCG in WGs.
From the nonlinear response of the dispersion-engineered WGs, we can infer information about the nonlinear optical parameters of the material, the quality of the epilayers, and fabrication process.
The attenuation of the nonlinearly generated signal in different frequency bands can show the influence of wavelength-dependent losses related to the presence of voids, sidewalls, and impurity-induced scattering, as well as cladding absorption.  
To this end, we tested the WGs fabricated from samples B and C following the fabrication process introduced in Sec. 2.2 for MRRs. 
The WG length was set at 4.5~mm, the height was fixed by the epilayer thickness, and the width was varied to tailor the modal dispersion.

SHG can occur only with a TM-polarized pump that aligns with the strongest component of the AlN nonlinear susceptibility tensor\cite{chen_calculated_1995,pernice_second_2012}. Figure ~\ref{fig:Nonlinear}(a) reports efficient SHG in 1.25~µm-wide WGs from the hybrid sample C while sweeping the TM pump in the telecom range. The peak corresponds to the intermodal phase-matching (PM) between the TM$_{00}$ pump and the TM\textsubscript{20} mode at the second harmonic (SH), and its position is strongly dependent on the exact WG geometry, as we show in the mode simulations reported in SI Sec.~S5. Moreover, the SH peak position can be finely tuned by varying the sample temperature. Indeed, as shown in Figure ~\ref{fig:Nonlinear}(a), by changing the sample temperature from 20$^\circ$C (T1) to 83$^\circ$C (T2), the SH peak shifts to longer wavelengths.

We then tested SCG through soliton dynamics and dispersive wave (DW) generation obtained by pumping the WGs with a femtosecond laser in the anomalous dispersion region, i.e. where $\beta_2<0$, with $\beta_2$ the second derivative of the wave number $\beta$ with respect to the angular frequency $\omega$. The pump has a central wavelength of 1560~nm, TM polarization, a pulse duration of 46~fs, and a repetition rate of 100~MHz. Light was coupled into the chip and collected at the output through an aspheric and an achromatic lens, respectively, and sent to optical spectrum analyzers covering the spectral range from the visible to the mid-IR.
The resulting spectra for 1.0~µm- and 1.3~µm-wide WGs are shown in the top panels of Figure~\ref{fig:Nonlinear}(b), for samples B (left) and C (right).
These are consistent with results obtained previously with similar geometries and power conditions \cite{lu_ultraviolet_2020}.
The supercontinuum (SC) coverage is primarily limited by the spectral positions of the long-wavelength DW (LWDW) and the short-wavelength DW (SWDW), located at phase-matched wavelengths between the solitons and the linear waves in the normal dispersion regime ($\beta_2>0$).
This PM condition depends on the WG cross-section, leading to a shift of the DW when WGs with the same height but different widths (same epilayer) or with the same width but different heights (different epilayers) are compared.
The bottom panels of Figures~\ref{fig:Nonlinear}(b) show the corresponding integrated dispersion, given by $\beta_{\text{int}} = \beta(\omega_{\text{DW}})-\beta(\omega_{\text{s}})-(\omega_{\text{DW}}-\omega_{\text{s}})/v_{\text{g,s}}$, with $\omega_{\text{DW}}$ and $\omega_{\text{s}}$ the angular frequency of the DW and soliton, and $v_{\text{g,s}}$ the soliton group velocity.
Applying the $\beta_{\text{int}}=0$ condition, we can find the DW position, indicated on the plots by yellow dots.
The good correspondence between the experimental and simulated positions of the DWs validates the chromatic dispersion profile that was considered and provides valuable insight into the precision of the WG dimension control. The SC extends further in the short wavelength side in sample C, compared to sample B, showing promising properties for visible light applications. The dispersion of the WGs of sample C, which are slightly thicker than those made from sample B, is also more favorable for pushing the LWDW deeper into the mid-IR.
However, our simulations indicate that the larger thickness of sample C relative to sample B would lead to a lower DW generation efficiency, as a result of significantly larger $|\beta_{\text{int}}|$ values and a reduced spectral overlap between the soliton and its corresponding DW. Experimentally we also see a reduction in the LWDW power, dominated by the above-mentioned effects. Optimization of the generation efficiency in thicker layers would require additional tuning of the pump wavelength, as shown in silicon nitride \cite{grassani_mid_2019, tagkoudi_parallel_2020} 
and AlN \cite{sbarra_three-octave_2025} WGs.  

Overall, these results demonstrate an efficient nonlinear response, control over the WG dispersion and dimensions, and confirm the suitability of the present hybrid AlN-on-sapphire epilayers as an integrated platform for harmonic and broadband light generation. The low transmission losses achieved through optimized growth and nanofabrication process, together with the ability of engineering the WG dispersion by adjusting their dimension, suggest the possibility of pushing the spectral coverage deeper into both the UV and mid-IR.

\section{Conclusion}

In conclusion, we experimentally showed the detrimental impact of voids on the optical performance of AlN-on-sapphire WGs and MRRs operating at \SI{1550}{\nano\meter}. Complementary light propagation FDTD simulations done for AlN straight WGs with and without voids unambiguously show their key role on scattering losses, which are mostly dependent on their size and density. Moreover, our simulation results indicate that even small voids, which are expected to cause negligible propagation losses at telecom wavelengths, have a noticeable impact on light propagation at shorter wavelengths, suggesting that voids may represent one of the potential limiting factors in current state-of-the art AlN photonics in the visible range.

Finally, we showed that a hybrid approach consisting of an MOVPE-grown AlN layer deposited on a thin sputtered AlN buffer leads to void-free epilayers characterized by low propagation losses in the telecom C-band, as confirmed by $Q_{\text{int}}$ values up to 2 million measured at \SI{1550}{\nano\meter} in MRRs. Such hybrid AlN epilayers also proved suitable for high-power nonlinear optical applications, as verified by SHG and broadband SCG measurements. Furthermore, since void-related scattering is expected to be more impactful at short wavelengths, the void-free nature of these hybrid AlN epilayers makes them appealing candidates for low-loss PICs operating in the visible and UV ranges, with the potential to advance the state-of-the-art performance of this platform.

\section{Method \label{par:method}}

\subsection{Growth of AlN epilayers} Sample A was grown in an Aixtron 200/4 RF-S horizontal reactor on a \textit{c}-plane single-side polished sapphire substrate with a miscut angle of \ang{0.2} toward the $m$-axis. A \SI{3}{\nano\meter}-thick AlN nucleation layer was first grown at a temperature of \SI{700}{\celsius} to initiate the Al polarity. AlN growth was then performed at a temperature of \SI{1100}{\celsius}, up to a final thickness of approximately \SI{1.1}{\micro\meter}. Sample B is a \SI{1.0}{\micro\meter}-thick commercially available epilayer on \textit{c}-plane single-side polished sapphire from DOWA Electronics Materials Co. Ltd. For sample C, a 50-nm-thick AlN buffer layer was first deposited via sputtering by Evatec AG onto a sapphire substrate identical to that of sample A.  MOVPE regrowth was then performed using the same reactor and the same growth procedure at \SI{1100}{\celsius} as for sample A, up to a similar total thickness ($\sim$\SI{1.1}{\micro\meter}).

\subsection{Device fabrication} Waveguiding structures were patterned onto the AlN epilayers using a 100 kV EBL machine (Raith EBPG5000+) and the negative-tone FOx-16 hydrogen silsesquioxane (HSQ) resist. A thin titanium (Ti) layer was evaporated onto the samples prior to resist coating to facilitate charge dissipation. The pattern was transferred from the FOx-16 resist to the AlN layers by means of a Cl\textsubscript{2}/BCl\textsubscript{3}/Ar-based ICP dry-etching process, with an AlN:FOx-16 etching selectivity of 2.2:1. After etching, the Ti and HSQ residues were removed with a buffered oxide etchant solution and the AlN structures were then cladded with \SI{3}{\micro\meter} of PECVD-deposited SiO\textsubscript{2}. Lastly, the chips were cleaved along the AlN \textit{m}-plane for edge coupling using the dice-and-cleave technique \cite{chen_fabrication_2014}. For simple WGs in sample A (Figure~\ref{fig:ring_Q}(a)), facet polishing was also employed to further even out coupling losses across the chip.

\subsection{Device characterization} The fabricated PIC devices were tested with a Toptica CTL 1550 tunable laser. Light was coupled to the chip with a microlensed fiber (spot size = \SI{2.3}{\micro\meter}) and collected either with a second microlensed fiber (when measuring the WG losses with the cut-back method) or with a free-space objective (for MRR characterization) and sent to an infrared photodetector. For samples B and C, given the narrow MRR resonance linewidths, the wavelength scan was performed through a voltage sweep on the laser piezoelectric actuators, calibrated with an electro-optical modulator and a signal generator in the GHz range.

\begin{acknowledgement}

This work was supported by the EPFL Science Seed Fund “Aluminum nitride-on-sapphire for nonlinear integrated photonics” and the Swiss National Science Foundation (SNSF) project 200020\_215633. We acknowledge Evatec AG for providing us with thin sputter-deposited AlN buffer layers on sapphire. 

\end{acknowledgement}

\begin{suppinfo}

Additional material characterization of the AlN epilayers used for this work including the formation of air pockets subsequent to PECVD SiO$_{2}$ cladding deposition. Additional details related to both sidewall scattering losses occurring in MRRs and FDTD simulations of void-induced scattering losses. Simulations of phase matching for SHG.

\end{suppinfo}

\bibliography{references}

\end{document}


\title{Supplementary Information for: \\ Sputtered AlN buffer layer for low-loss crystalline AlN-on-sapphire integrated photonics}
\author{
    Samuele Brunetta$^{*,\dag}$, Samantha Sbarra$^{\ddag}$, Brandon Shuen Yi Loke$^{\ddag}$, \\ Jean-François Carlin$^{\dag}$, Nicolas Grandjean$^{\dag}$, Camille-Sophie Brès$^{\ddag}$, and \\ Raphaël Butté$^{\dag}$
}
\date{}
\maketitle

\begin{center}
$^{\dag}$Laboratory of Advanced Semiconductors for Photonics and Electronics, Ecole Polytechnique Fédérale de Lausanne (EPFL), CH-1015 Lausanne, Switzerland \\
$^{\ddag}$Photonic Systems Laboratory, Ecole Polytechnique Fédérale de Lausanne (EPFL), CH-1015 Lausanne, Switzerland \\
$^{*}$Corresponding author: \texttt{samuele.brunetta@epfl.ch}
\end{center}

\tableofcontents
\newpage

\section{Additional material characterization of AlN-on-sapphire epilayers}

In this section we provide additional material characterization data for the three aluminum nitride (AlN) epilayers used in this work. The samples were all grown on 2-inch \textit{c}-plane sapphire substrates by metalorganic vapor-phase epitaxy (MOVPE). Sample A was grown in house using a full-MOVPE process, which includes a thin AlN nucleation layer grown at lower temperature to facilitate the start of the heteroepitaxial growth process. Sample B is a \SI{1.0}{\micro\meter}-thick commercially available epilayer purchased from DOWA Electronics Materials Co. Ltd. For sample C, a hybrid growth process was used, which consists of the MOVPE growth of an approximately \SI{1.0}{\micro\meter}-thick AlN layer on a 50-nm-thick sputter-deposited AlN buffer layer on \textit{c}-plane sapphire provided by Evatec AG. Additional details are provided in the Method section of the main text. 

\begin{figure}[ht]
   \begin{center}
   \includegraphics[width=1.0\textwidth]{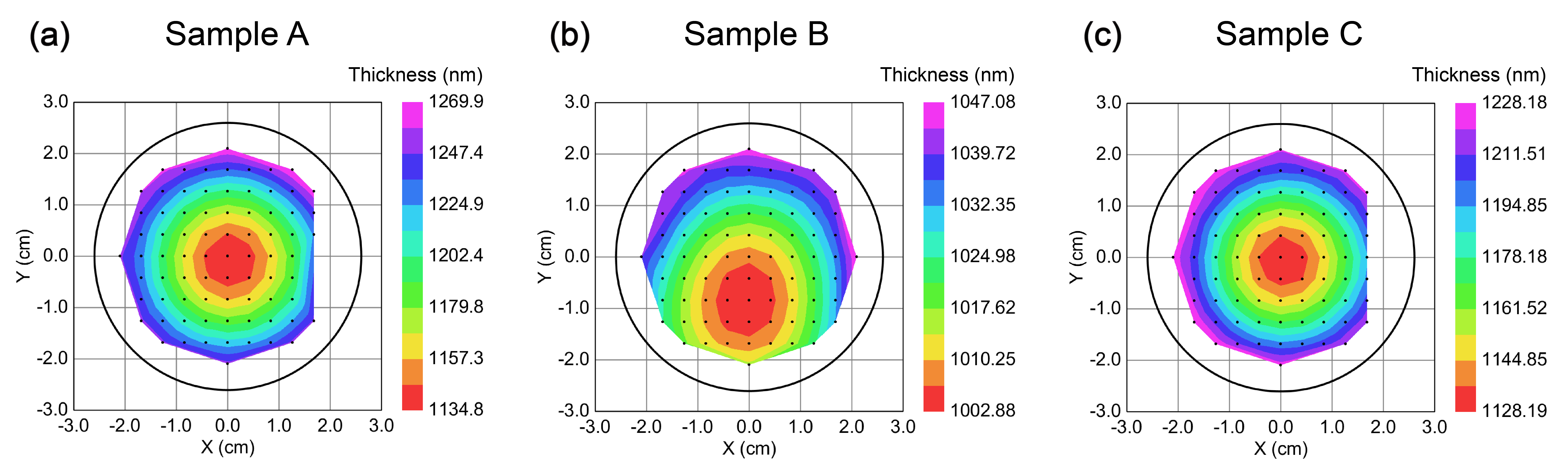}
   \end{center}
   \caption{(a) to (c) 2D thickness maps of  epilayers A, B and C, respectively, measured by spectroscopic ellipsometry.}
  \label{fig:ellipsometry_SM}
   \end{figure} 

\begin{figure}[ht]
   \begin{center}
   \includegraphics[width=1.0\textwidth]{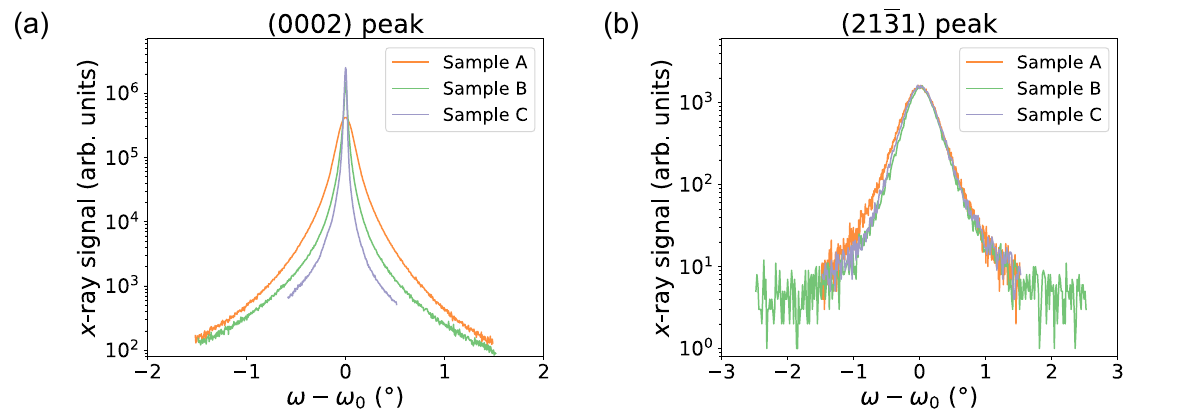}
   \end{center}
   \caption{XRD rocking curves of the three AlN epilayers under consideration. (a) Results for the ($0002$) symmetric peak. (b) Results for the ($21\bar{3}1)$ asymmetric peak.}
  \label{fig:XRD_SM}
   \end{figure}

In Figures~\ref{fig:ellipsometry_SM}(a) to \ref{fig:ellipsometry_SM}(c), we show the two-dimensional (2D) AlN thickness maps of epilayers A, B and C, respectively, obtained by spectroscopic ellipsometry using a Woollam RC2 tool. Single-point measurements at the wafer center, performed in the 300-2500~nm wavelength range and at incident angles of \ang{55}, \ang{65} and \ang{75}, were used to determine the refractive index curve of the layer by fitting the data with a uniaxial Sellmeier model. Then, scans at a \ang{55} angle were taken across the whole wafer in order to extract the thickness map using the refractive index data obtained with the single-point measurements. In all the samples, the AlN layer  thickness increases from the wafer center toward the edges. However, while the thickness variation across the wafer is about \SI{100}{\nano\meter} for the in-house grown samples A and C, the commercially available sample B exhibits thickness variations less than \SI{50}{\nano\meter}. These features are most likely stemming from differences in the growth parameters and the reactor geometries (single 2-inch wafer capability for the in-house horizontal reactor). Thickness variation can be a critical parameter in applications that require precise dispersion engineering, such as second-harmonic generation and supercontinuum generation. In contrast, it plays a much less significant role when the objective consists in extracting propagation losses from waveguides (WGs) or microring resonators (MRRs). Moreover, for individual devices fabricated from the full wafer (e.g., our chips typically have dimensions of $4.5\times \SI{10}{\square\milli\meter}$), the local thickness variation is much smaller, and the design of photonic components---such as WGs---can be adjusted on each single chip based on the 2D thickness map of the epilayer.

X-ray diffraction (XRD) data for the  (0002) symmetric and ($21\bar{3}1)$ asymmetric peaks are shown in Figures~\ref{fig:XRD_SM}(a) and \ref{fig:XRD_SM}(b), respectively. While the three epilayers show little difference in the asymmetric peak, sample A has a broader symmetric peak than samples B and C, suggesting that sample A has a poorer crystalline quality and a higher degree of off-plane crystal orientation than both samples B and C (which are instead comparable to each other). The values of the full width at half maximum (FWHM) for both peaks in the three samples are those reported in Table~1 of the main text.

\begin{figure}
   \begin{center}
   \includegraphics[width=1.0\textwidth]{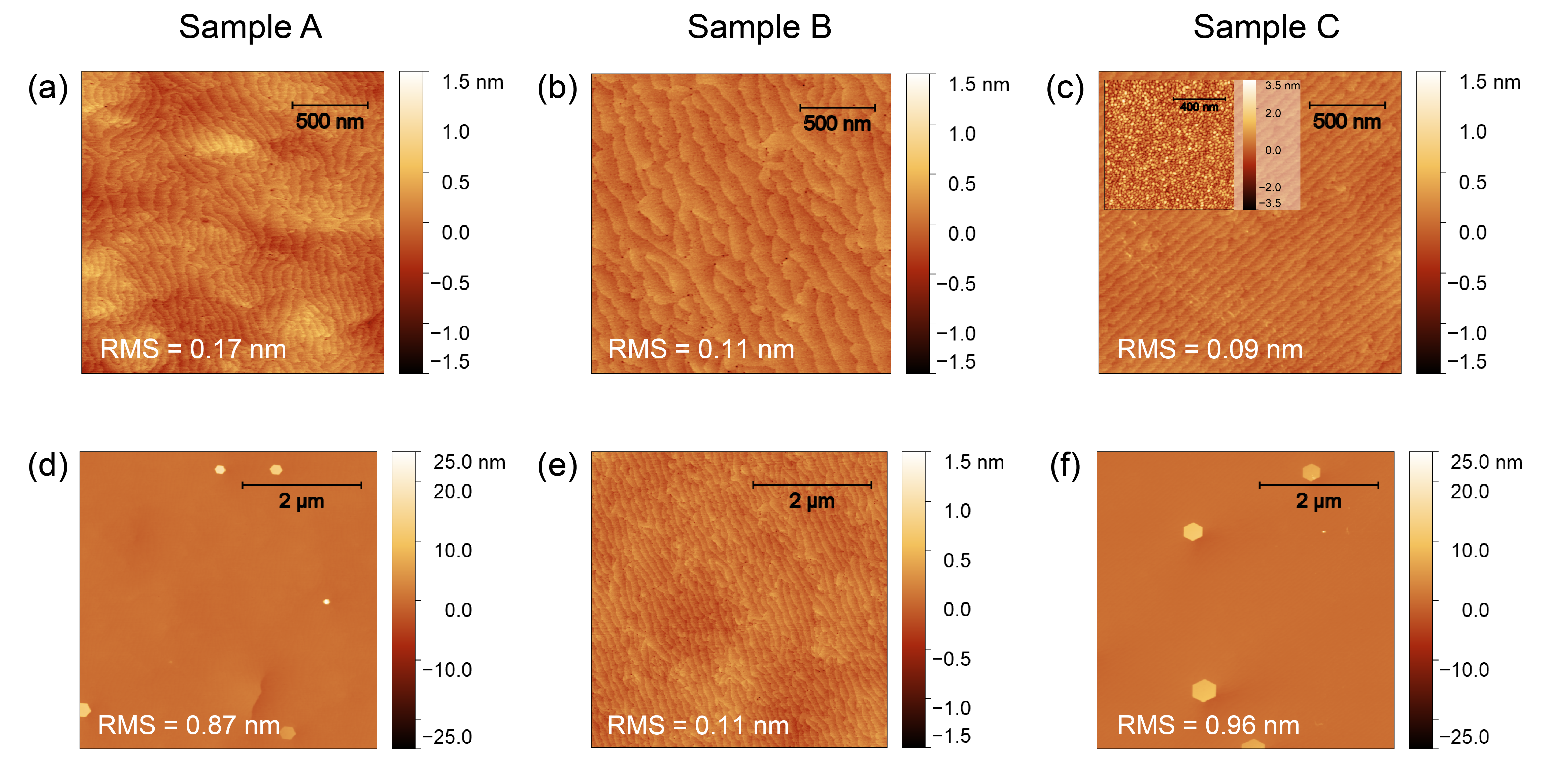}
   \end{center}
   \caption{(a) to (c) $2\times \SI{2}{\square\micro\meter}$ AFM surface scans of epilayers A, B and C, respectively, with the extracted value of the RMS surface roughness. The inset in Figure~\ref{fig:AFM_SM}(c) shows a $1\times \SI{1}{\square\micro\meter}$ surface scan of the \SI{50}{\nano\meter} thick AlN sputtered buffer layer prior to MOVPE regrowth, yielding an RMS surface roughness of \SI{0.72}{\nano\meter}.        (d) to (f) $5\times \SI{5}{\square\micro\meter}$ AFM surface scans of epilayers A, B and C, respectively, revealing the presence of hexagonal hillocks and faceted macrosteps in samples A and C.}
  \label{fig:AFM_SM}
   \end{figure}    

Surface roughness and topography were probed by means of atomic force microscopy (AFM). In Figures~\ref{fig:AFM_SM}(a), \ref{fig:AFM_SM}(b) and \ref{fig:AFM_SM}(c), we report $2\times \SI{2}{\square\micro\meter}$ AFM surface scans of samples A, B and C, respectively, whose extracted value of the root mean square (RMS) surface roughness remains lower than \SI{0.2}{\nano\meter} for all the samples. In the inset of Figure~\ref{fig:AFM_SM}(c), we show a $1\times \SI{1}{\square\micro\meter}$ map of the \SI{50}{nm}-thick sputtered AlN buffer layer prior to MOVPE regrowth. The layer has an RMS surface roughness of \SI{0.72}{\nano\meter}, which is higher than the value of \SI{0.09}{\nano\meter} measured for the full layer after regrowth. This means that the MOVPE growth process manages to recover the rougher surface of the sputtered layer and produces a remarkably smooth surface in the final layer.

Although all the samples present smooth surfaces at small scale, larger AFM scans such as shown in Figures~\ref{fig:AFM_SM}(d) to \ref{fig:AFM_SM}(f) reveal the presence of some hexagonal hillocks and faceted macrosteps in samples A and C. These hillocks, which are not found in the commercially available sample B, are present with densities of \SI{1.1e7}{\per\square\centi\meter} in sample A and \SI{2.1e7}{\per\square\centi\meter} in sample C, and their height ranges between 5 and \SI{30}{\nano\meter}. Taking hillocks into account, the RMS surface roughness of samples A and C is higher, as shown in Figures~\ref{fig:AFM_SM}(d) and \ref{fig:AFM_SM}(f), but still below the \SI{1}{\nano\meter} threshold. Hillocks in MOVPE grown AlN are commonly associated with several factors including growth temperature, the presence of $\gamma$-AlON islands, the V/III ratio and surface preparation.\cite{peters_parasitic_2023, pampili_nitrogen-polar_2024} However, optimizing the growth recipe can be a complex and time-consuming process and the presence of hillocks did not visibly impact propagation losses at \SI{1550}{\nano\meter} in our devices, as we were able to fabricate MRRs from sample C with an intrinsic quality factor ($Q_{\text{int}}$) up to $\SI{2.0d6}{}$, corresponding to optical losses of \SI{0.19}{\decibel\per\centi\meter}. Therefore, we decided not to focus our attention on minimizing or even suppressing hillocks in this work. However, hillocks could potentially cause some non-negligible surface scattering losses at shorter wavelengths. Hence, their presence should likely be minimized in AlN epilayers intended for such applications. 

   \begin{figure}
   \begin{center}
   \includegraphics[width=1.0\textwidth]{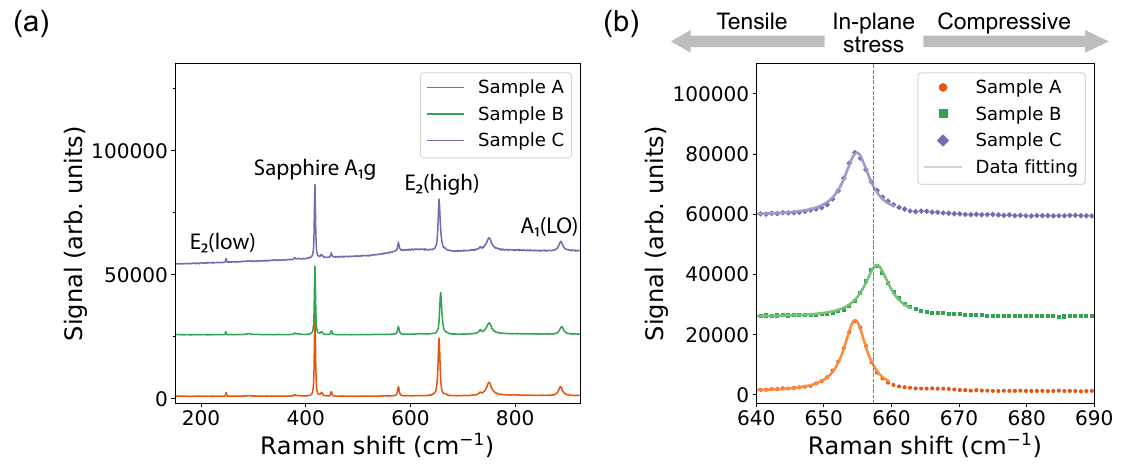}
   \end{center}
   \caption{(a) Raman spectra of the three epilayers recorded using cw \SI{532}{\nano\meter} laser excitation. (b) Raman data for the AlN E\textsubscript{2}(high) mode, with their corresponding Lorentzian fit, showing different stress conditions in the three samples. }
  \label{fig:Raman_SM}
   \end{figure} 

Biaxial stress was estimated by means of micro-Raman spectroscopy using a Renishaw inVia Reflex Raman Confocal microscope. In Figure~\ref{fig:Raman_SM}(a), we show Raman spectra recorded on the three epilayers using continuous wave (cw) \SI{532}{\nano\meter} laser excitation. The zoomed-in results for the AlN E\textsubscript{2}(high) mode are shown in Figure~\ref{fig:Raman_SM}(b). By fitting the data with a Lorentzian curve we extracted an E\textsubscript{2}(high) peak center $\omega$ of \SI{654.6}{\per\centi\meter}, \SI{657.8}{\per\centi\meter} and \SI{654.9}{\per\centi\meter} for samples A, B and C, respectively. The biaxial stress $\sigma_{xx}$ can be calculated using the relationship:\cite{he_fast_2020}
\begin{equation}
    \sigma_{xx}=\frac{\omega-\omega_0}{k},
    \label{eq:stress}
\end{equation}
where $\omega_0= \SI{657.4}{\per\centi\meter}$ is the E\textsubscript{2}(high) peak position of strain-free aluminum nitride\cite{davydov_phonon_1998} and $k=\SI{-4.04}{\per\centi\meter\per\giga\pascal}$ is the biaxial stress coefficient.\cite{yang_raman_2011} Using Eq.~\eqref{eq:stress} we obtain values of $\sigma_{xx}$ of $\SI{0.69}{\giga\pascal}$, $\SI{-0.11}{\giga\pascal}$ and $\SI{0.62}{\giga\pascal}$ for samples A, B and C, respectively. These results indicate that, while the commercially available sample B is under slight compressive stress, samples A and C are subjected to tensile stress, which could likely explain the presence of a few isolated cracks in both samples A and C.

\section{Bus-microring gap filling capability of silane-based PECVD}

\begin{figure}[ht]
   \begin{center}
   \includegraphics[width=1.0\textwidth]{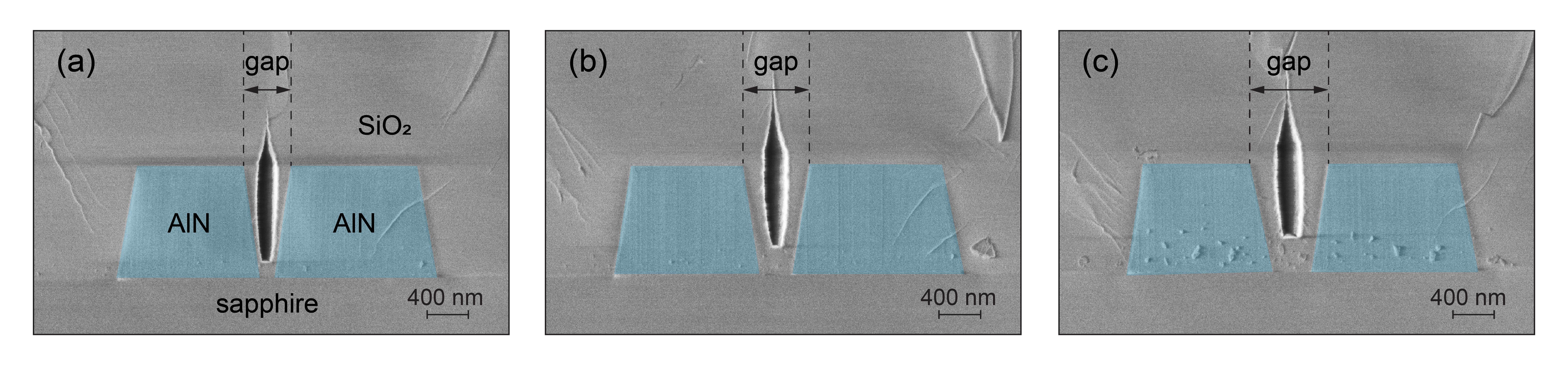}
   \end{center}
   \caption{Cross-section SEM images showing the PECVD SiO\textsubscript{2} trench filling capability for different bus-microring gap configurations on sample B. (a) $\mathrm{gap}=\SI{500}{\nano\meter}$. (b) $\mathrm{gap}=\SI{700}{\nano\meter}$. (c) $\mathrm{gap}=\SI{800}{\nano\meter}$.}
   \label{fig:SEM_gap_SM}
   \end{figure}

As mentioned in the main text, the SiO\textsubscript{2} cladding layer was deposited by silane-based  plasma-enhanced chemical vapor deposition (PECVD). This technique is known to leave air pockets when filling narrow trenches such as the bus-microring gap. Since most of the undercoupled MRRs we characterized in samples B and C have gaps ranging between \SI{500}{} and \SI{800}{\nano\meter}, we report in Figure~\ref{fig:SEM_gap_SM} false-color cross-section scanning electron microscopy (SEM) images showing the PECVD trench filling capability in this range of gaps. As can be observed, an air pocket of non-negligible size is always present, even for the largest gap we considered. This air pocket can perturb the bus-microring coupling and act as a source of scattering. Therefore, it is probably one of the main factors currently limiting the performance of the fabricated devices. Potential improvements could be obtained by using tetraethyl-orthosilicate-based PECVD, which has superior trench refilling capability compared to silane-based PECVD.\cite{chang_trench_2004, sun_ultrahigh_2019} However, this aspect goes beyond the scope of this work and is left for future investigations.

\section{Evaluation of sidewall scattering losses in microring resonators}

\begin{figure}[ht]
   \begin{center}
   \includegraphics[width=1.0\textwidth]{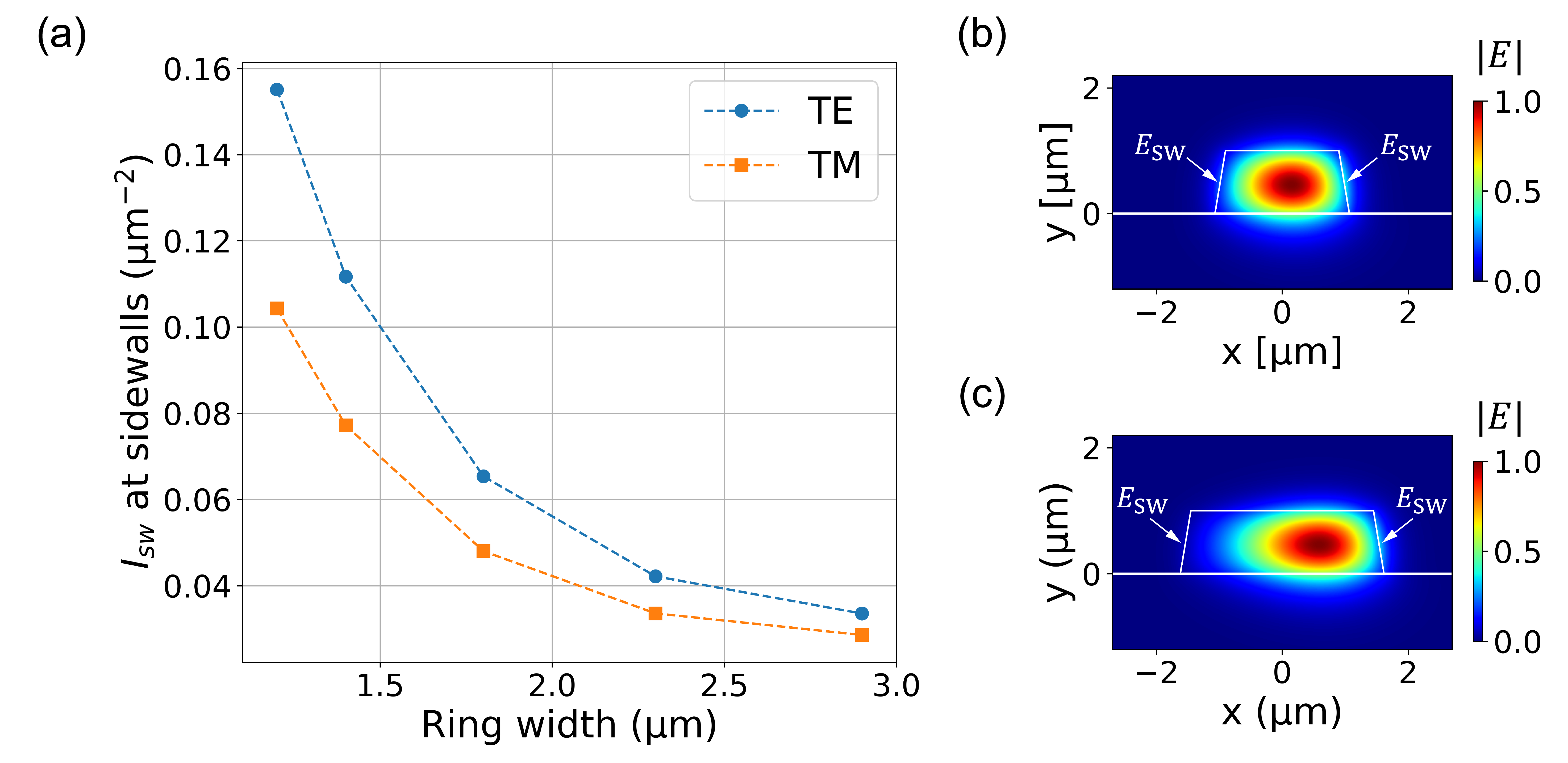}
   \end{center}
   \caption{(a) Normalized field intensity $I_\mathrm{sw}$ at the sidewalls for MRRs of \SI{60}{\micro\meter} radius and different ring widths. (b) and (c) Normalized  TE\textsubscript{00} field maps at $\lambda=$\SI{1550}{\nano\meter} for $\SI{1.8}{\micro\meter}$- and $\SI{2.9}{\micro\meter}$-wide MRRs, respectively, showing a reduced amount of field in the inner sidewall for the latter. The asymmetric mode shape is due to the radius of curvature of the MRRs.}
  \label{fig:sidewall_field_SM}
   \end{figure} 

In order to identify to which extent sidewall roughness is a limiting factor in the performance of our MRRs, we followed the same analytical approach as that described in Ref.~\citenum{liu_ultra-high-q_2018}. In optical WGs, sidewall-related scattering losses $\alpha_\mathrm{sw}$ at fixed wavelength are proportional to:\cite{deri_low-loss_1991, payne_theoretical_1994}
\begin{equation}
    \alpha_\mathrm{sw} \propto \frac{\sigma^2 (n_\mathrm{core}^2-n_\mathrm{clad}^2)^2}{ n_\mathrm{eff}} \frac{ \frac{1}{l} \int E_\mathrm{sw}^2 \, \mathrm{d}l}{\iint E^2 \, \mathrm{d}x \, \mathrm{d}y},
    \label{eq:roughness_SM}
\end{equation}
where $\sigma$ is the sidewall roughness, $n_\mathrm{core}$ and $n_\mathrm{clad}$ are the refractive indices of the core and cladding materials, respectively, and $n_\mathrm{eff}$ is the effective index of the mode. For the same level of surface roughness and approximating $n_\mathrm{eff}$ as a constant, we infer from Eq.~\eqref{eq:roughness_SM} that $\alpha_\mathrm{sw}$ is proportional to the normalized average field intensity at the sidewalls $I_\mathrm{sw} \equiv (\frac{1}{l} \int E_\mathrm{sw}^2 \, \mathrm{d}l)/\iint E^2 \, \mathrm{d}x \, \mathrm{d}y$.

Through a series of finite-difference-eigenmode (FDE) simulations performed with Ansys Lumerical, we extracted the mode profile of all the MRRs studied in Figure~5 of the main text, and for each of them we calculated the normalized field intensity at the sidewall $I_\mathrm{sw}$, for both the fundamental TE and TM modes. Results are shown in Figure~\ref{fig:sidewall_field_SM}(a), in which we can spot a decrease of $I_\mathrm{sw}$ with increasing ring width. This is due to the fact that wider ring wings experience a lower amount of sidewall field $E_\mathrm{sw}$, as also noticeable from the mode profiles shown in Figures~\ref{fig:sidewall_field_SM}(b) and \ref{fig:sidewall_field_SM}(c). 

In particular, from Figure~\ref{fig:sidewall_field_SM} we observe that for a ring width of \SI{2.9}{\micro\meter}, $I_\mathrm{sw}$ is about half the value computed for a ring width of \SI{1.8}{\micro\meter}. This means that, if propagation losses were only due to sidewall scattering, we should measure a  $Q_{\text{int}}$ value approximately twice larger for a ring width of \SI{2.9}{\micro\meter} compared to \SI{1.8}{\micro\meter} for both TE and TM polarizations. However, this is not in agreement with the experimental results obtained with sample B shown in Figure~5 of the main text, for which $Q_{\text{int}}$ values for widths larger than \SI{1.8}{\micro\meter} reach a plateau. This allows us to conclude that sidewall scattering is not the main factor limiting propagation losses for our wider MRRs. Intrinsic quality factors in our samples are therefore likely limited by other influences, such as the air pocket in the cladding at the bus-microring gap and the presence of PECVD-related O-H bonds in the cladding, as we also discuss in the main text.

\section{3D-FDTD simulation settings and convergence}

\begin{figure}[ht]
   \begin{center}
   \includegraphics[width=1.0\textwidth]{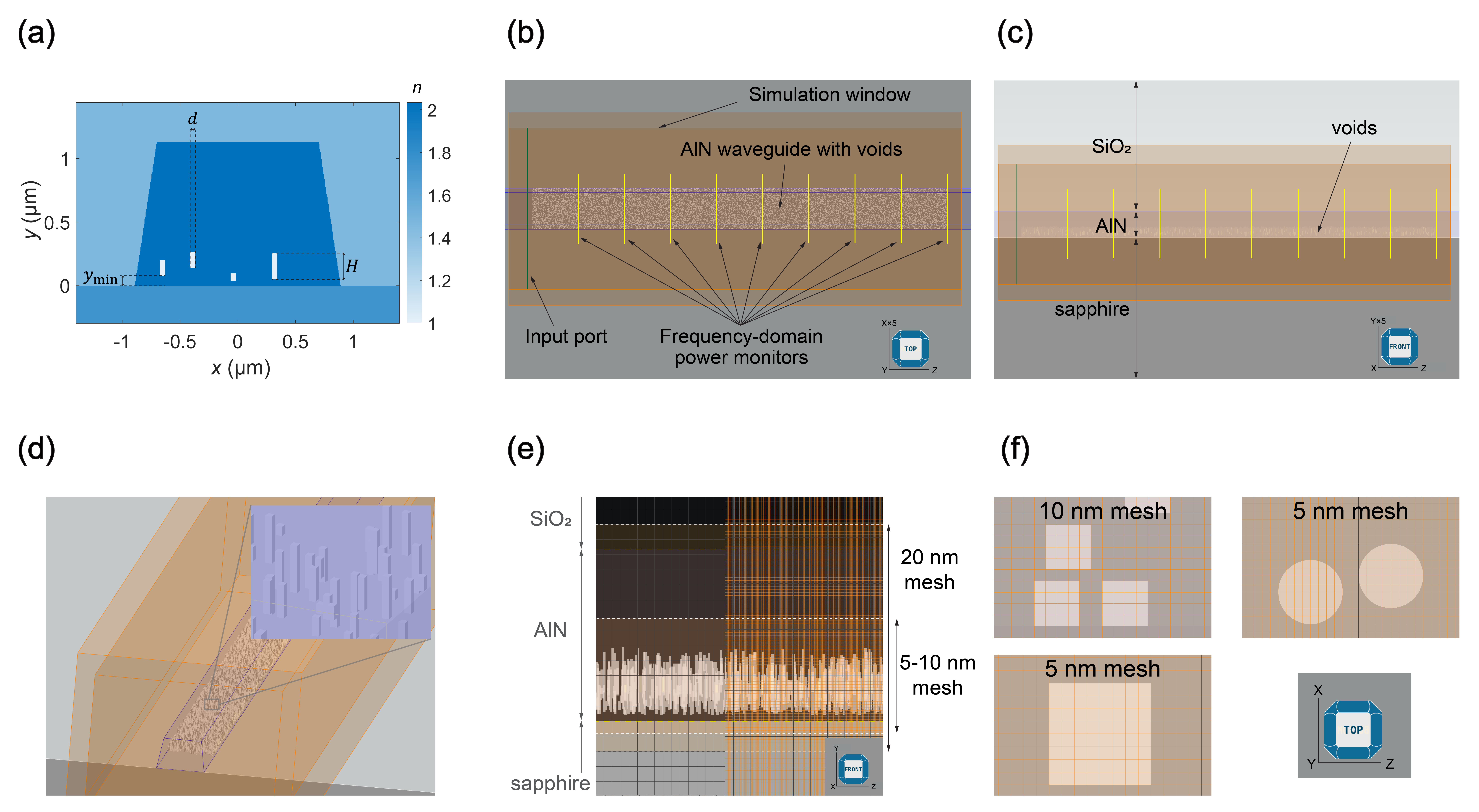}
   \end{center}
   \caption{(a) Cross-section refractive index map of the simulated WG showing the presence of voids and the definition of relevant geometric parameters. (b) and (c) 3D-FDTD simulation setting schemes for the $(x,z)$ and $(y,z)$ planes, respectively. (d) 3D representation of parallelepiped-shaped voids inside a WG including a higher magnification inset. (e) $(y,z)$ plane scheme showing the different meshes implemented in the simulations. (f) $(x,z)$ meshing of voids with different geometries and mesh sizes.}
  \label{fig:simulations_SM}
   \end{figure}

In this section, we provide additional details on the finite-difference time-domain (FDTD) simulations of void-related propagation losses discussed in the main text. Simulations were performed with the three-dimensional (3D) FDTD module of Ansys Lumerical.\footnote{Ansys Lumerical FDTD v. 8.30.3536 from Lumerical Inc.} We simulated light propagation in the telecom C-band of an AlN-on-sapphire WG (height: \SI{1.1}{\micro\meter}, upper width: \SI{1.4}{\micro\meter}) having voids. Voids were modeled as parallelepipeds with height $H$, in-plane diameter $d$, bottom coordinate $y_\mathrm{min}$ and areal density $N$. The definition of these quantities is also shown graphically on the refractive index map of Figure~\ref{fig:simulations_SM}.  The extent of the simulation region was set to \SI{7}{\micro\meter} in width and \SI{5.2}{\micro\meter} in height, and perfectly matched layer boundary conditions were adopted. The mesh refinement method was kept in the ``conformal variant 0" default option, which is the most consistent scheme as advised by the Lumerical support service. The fundamental TE and TM modes were injected through input ports placed in a void-free region of the WG. The total WG region under the influence of voids was chosen to be \SI{90}{\micro\meter} long, and frequency-domain power monitors were placed at a pitch of \SI{10}{\micro\meter}, as shown in the top-view and lateral-view schemes of Figures~\ref{fig:simulations_SM}(b) and \ref{fig:simulations_SM}(c). A fine simulation mesh of \SI{10}{\nano\meter} was used in the bottom part of the WG, where voids are present. In the rest of the WG region---including at least \SI{160}{\nano\meter} outside the WG---a \SI{20}{\nano\meter} mesh was employed, while a coarser mesh was used in the remaining parts of the simulation domain. A lateral-view scheme picturing the various simulation meshes is shown in Figure~\ref{fig:simulations_SM}(e).
\begin{table}[t]
\small
\caption{Convergence studies of 3D-FDTD simulation results consisting of analyzing optical losses for different mesh sizes. The effect of void shape is also taken into consideration.}
  \label{tbl:sim_data_SM}
  \centering
\begin{tabular}{cccccc}
\hline 
Void shape     & $d$  & $xz$ area (nm\textsuperscript{2}) & Mesh size (nm) & \begin{tabular}[c]{@{}c@{}}TE\\ $\alpha$ (\SI{}{\decibel\per\centi\meter})\end{tabular} & \begin{tabular}[c]{@{}c@{}}TM\\ $\alpha$ (\SI{}{\decibel\per\centi\meter})\end{tabular} \\ \hline
Parallelepiped & 30 & 900          & 10             & $8.2\pm0.1$                                    & $5.1\pm0.1$                                    \\
Parallelepiped & 40 & 1600         & 10             & $21.7\pm0.6$                                  & $16.4\pm0.3$                                   \\
Parallelepiped & 40 & 1600         & 5              & $22.8\pm0.8$                                   & $16.8\pm0.4$                                   \\
Cylinder       & 40 & 1257         & 5              & $15.0\pm0.6$                                   & $10.6\pm0.3$  \\
\hline
\end{tabular}
\end{table}

Several tests were carried out to check the reliability and convergence of our simulation settings.  The main results are reported in Table~\ref{tbl:sim_data_SM}. For these tests, the following void parameters were adopted: $H$ and $y_\mathrm{min}$ were randomly distributed and both lied in the 40-200~nm range while $N$ and $d$ were kept fixed at $\SI{100}{\per\square\micro\meter}$ and $\SI{40}{\nano\meter}$, respectively. A convergence test was performed by reducing the mesh size in the void region from \SI{10}{} to \SI{5}{\nano\meter}. In this latter case, the WG length under the influence of voids was shortened to \SI{60}{\micro\meter} to reduce the computational load. Losses for both TE and TM polarizations were calculated using the transmission results from the power monitors with the method detailed in the main text. A comparison of the second and third rows of Table~\ref{tbl:sim_data_SM} shows that propagation losses vary by less than \SI{5}{\percent} for the \SI{10}{} and \SI{5}{\nano\meter} meshes. Therefore, since computational times were about 5 times shorter with the \SI{10}{\nano\meter} mesh, we adopted this configuration for all the simulations shown in the main text. 

In addition, we also studied the effect of void geometry by simulating propagation losses for cylindrical voids with a diameter of \SI{40}{\nano\meter}. In this case, a \SI{5}{\nano\meter} mesh was adopted in order to better describe the round shape geometry of the voids. As can be seen from the last row of Table~\ref{tbl:sim_data_SM}, losses for cylindrical voids with $d=\SI{40}{\nano\meter}$ are in the range of values for parallelepiped voids with $d=\SI{30}{}$ and $\SI{40}{\nano\meter}$. From this table, we can notice that the in-plane area of cylindrical voids with $d=\SI{40}{\nano\meter}$ ranges also between those of parallelepiped voids with $d=\SI{30}{}$ and $\SI{40}{\nano\meter}$. This means that scattering is mainly dominated by the void size rather than by their exact in-plane geometry. The geometry still plays a role in the weight of propagation losses, but mainly in determining the ratio between TE and TM losses, as shown in the $H$ sweep results reported in Figure~6(f) of the main text.

Finally, we comment on the choice of using frequency-domain power monitors to extract transmission results. Power monitors have the peculiarity of collecting all the optical power passing through the monitor region. This may be advantageous, as they also collect the power that is temporarily stored outside the fundamental propagating mode (e.g., in higher-order modes). However, they also capture part of the optical power scattered out of the WG that still propagates within the solid angle defined by the monitor. Therefore, we chose the monitor size in order to collect the optical power of the propagating mode while minimizing the collection of the scattered field. Nevertheless, some scattered light is still captured by the monitors, hence contributing to the loss calculations. By performing a longer simulation where we placed additional monitors after the WG region occupied by voids, we concluded that the values of the propagation losses extracted from the 3D-FDTD simulations are likely underestimated by approximately \SI{8}{\percent}. Nonetheless, this remains the most reliable method we could find, given that monitoring only the optical power in the fundamental mode resulted in noticeable data oscillations, which made the loss estimation process more complicated and hence less reliable. 

\section{Simulations of phase matching for SHG}

In the main text, we show second-harmonic generation (SHG) from linear waveguides fabricated from sample C, pumped in the telecom range. To generate a second-harmonic (SH) signal, we targeted phase matching (PM) between the fundamental TM\textsubscript{00} mode and the SH TM\textsubscript{20} mode, as has been done in several previous studies on the same material platform.\cite{pernice_second_2012, guo_second-harmonic_2016, bruch_17_2018}

\begin{figure}[ht]
   \begin{center}
   \includegraphics[width=1.0\textwidth]{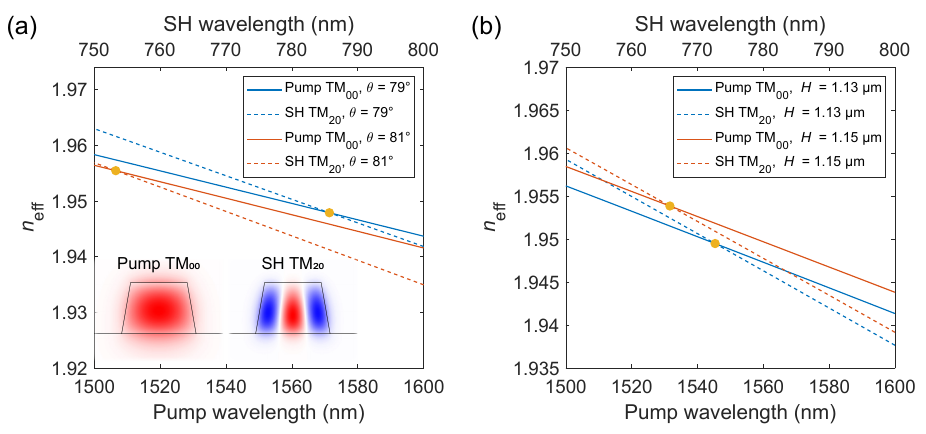}
   \end{center}
   \caption{(a) Simulations of the effective index dispersion of the pump TM\textsubscript{00} and the SH TM\textsubscript{20} modes for a WG geometry of $\SI{1.25}{}\times\SI{1.14}{\micro\meter}^{2}$ and varying sidewall angles $\theta$. The intersections between the two lines (yellow dots) correspond to the phase-matching conditions for the respective geometries. (b) Simulations for a WG width of \SI{1.25}{\micro\meter}, a sidewall angle of \ang{80}, and varying WG height $H$.}
  \label{fig:SHG_PM_SM}
   \end{figure}

To support our experimental results, we report mode simulations of AlN WGs, performed with the FDE module of Ansys Lumerical. In Figure~\ref{fig:SHG_PM_SM}(a), we show the effective index ($n_{\mathrm{eff}}$) dispersion of the two aforementioned modes in a WG with an upper width of $\SI{1.25}{\micro\meter}$, determined by lithography, and a height of $\SI{1.14}{\micro\meter}$, measured on the chip. Simulations show that varying the sidewall angle $\theta$ between
\ang{79} and \ang{81}, within our measurement uncertainty, consistently yields PM in the telecom range, although the PM wavelength depends strongly on the exact value of $\theta$. The experimentally observed peak position falls well within this uncertainty range, confirming a good agreement between simulations and experimental results.
A similar behavior for the WG height ($H$) can be observed in Figure~\ref{fig:SHG_PM_SM}(b), which shows that variations in $H$ within the experimental uncertainty, while keeping $\theta$ fixed at \ang{80}, also cause non-negligible shifts in the PM wavelength.

\bibliography{references}